\newif\ifsubmode
\submodefalse

\newif\ifprintfig
\printfigtrue

\newif\ifemulate
\emulatetrue

\ifsubmode
  \emulatefalse
  \documentclass[12pt,preprint]{aastex}
  \usepackage{epsfig}
  \received{}
  \accepted{}
  \journalid{}{}
  \articleid{}{}
\else
  \ifemulate
    \documentclass[apj]{emulateapj}
  \else
    \documentclass[preprint]{aastex}
  \fi
  \usepackage{epsfig}
  \slugcomment{{\it ApJ, submitted 11/17/2006}}
\fi

\shorttitle{Dynamical Models of Elliptical Galaxies in $z = 0.5$ clusters: II} 
\shortauthors{van der Marel \& van Dokkum}

\newcommand{\etal}{{et al.~}}
\newcommand{\lta}{\lesssim}
\newcommand{\gta}{\gtrsim}
\newcommand{\kms}{\>{\rm km}\,{\rm s}^{-1}}

\newcommand{\Mpc}{\>{\rm Mpc}}
\newcommand{\Msun}{\>{\rm M_{\odot}}}

\begin{document}

\title{Dynamical Models of Elliptical Galaxies in $z = 0.5$ Clusters:\\
II.~Mass-to-Light Ratio Evolution without Fundamental Plane Assumptions}

\author{Roeland P.~van der Marel}
\affil{Space Telescope Science Institute, 3700 San Martin Drive, 
       Baltimore, MD 21218}

\author{Pieter G.~van Dokkum}
\affil{Department of Astronomy, Yale University, New Haven, CT 06520}


\ifsubmode\else\ifemulate\else
\baselineskip=14pt
\fi\fi


\begin{abstract}
We study $M/L$ evolution of early-type galaxies using dynamical
modeling of resolved internal kinematics. This makes fewer assumptions
than Fundamental Plane (FP) studies and provides a powerful new
approach for studying galaxy evolution. We focus on the sample of 25
galaxies in clusters at $z \approx 0.5$ modeled in Paper~I. For
comparison we compile and homogenize $M/L$ literature data for 60
nearby galaxies that were modeled in comparable detail. The nearby
sample obeys $\log (M/L)_B = Z + S \log(\sigma_{\rm eff}/[200 \kms])$,
with $Z = 0.896 \pm 0.010$, $S = 0.992 \pm 0.054$, and $\sigma_{\rm
eff}$ the effective velocity dispersion. The $z
\approx 0.5$ sample follows a similar relation but with lower
zeropoint. The implied $M/L$ evolution is $\Delta \log(M/L) /
\Delta z = -0.457 \pm 0.046 \> {\rm (random)} \pm 0.078 \> {\rm
(systematic)}$, consistent with passive evolution following
high-redshift formation. This agrees with the FP results for this
sample by van Dokkum \& van der Marel. This confirms that FP evolution
tracks $M/L$ evolution, which is an important verification of the
assumptions that underly FP studies. However, while we find more FP
evolution for galaxies of low $\sigma_{\rm eff}$ (or low mass), the
dynamical $M/L$ evolution instead shows little trend with $\sigma_{\rm
eff}$. We argue that this difference can be plausibly attributed to a
combination of two effects: (a) evolution in structural galaxy
properties other than $M/L$; and (b) the neglect of rotational support
in studies of FP evolution. The results leave the question open
whether the low-mass galaxies in the sample have younger population
ages than the high-mass galaxies. This highlights the general
importance in the study of population ages for complementing dynamical
measurements with broad-band colors or spectroscopic population
diagnostics. 
\end{abstract}



\clearpage


\section{Introduction}
\label{s:intro}

A stellar population of fixed mass fades as it ages. The predicted
mass-to-light ratio $M/L$ of a galaxy therefore depends strongly on
its age (e.g., Bruzual \& Charlot 2003) and $M/L$ measurements can
constrain galaxy formation times. The $M/L$ of a galaxy cannot
generally be accurately constrained from characteristics of the
observed light (e.g., broad-band colors and line-strength indices)
alone. This is because low-mass stars contribute the bulk of the mass,
but only a small fraction of the light. Plausible variations in the
assumed initial mass function (IMF) can therefore change the $M/L$
without affecting significantly the characteristics of the observed
light (e.g., Bell \& de Jong 2001). Accurate measurements of galaxy
$M/L$ values therefore generally rely on dynamical properties. The
speed at which stars or gas move probes directly the gravitational
force to which they are subjected, and therefore the mass of the system.

The quality of the kinematical data for galaxies in the local
Universe, as well as the methods by which their dynamics can be
modeled, have steadily increased in sophistication over the years.
Large samples of reliable $M/L$ measurements are now available from
many studies, in particular for early-type galaxies (e.g., van der
Marel 1991; Magorrian \etal 1998; Gebhardt \etal 2003; Cappellari
\etal 2006a). For these galaxies it is well established that the $M/L$ 
inside an effective radius is relatively constant, and contains only a
small contribution from the dark halo (e.g., Kronawitter \etal 2000).
Therefore, the inferred $M/L$ values are primarily driven by the
characteristics of the stellar population. This situation differs
considerably from the case for spiral galaxies. While mass profiles
have been derived for many spiral galaxies (e.g., Persic, Salucci \&
Stel 1996), these profiles constrain primarily the properties of the
dark halo. Considerable uncertainty remains on the $M/L$ of the
stellar population, or alternatively, on whether spiral galaxy disks
are ``maximal'' or not (e.g., van Albada \etal 1985; Courteau \& Rix
1999).

Uncertainties in the IMF for low-mass stars cause considerable
uncertainty in the predicted $M/L$ at fixed age. So while accurate
dynamical $M/L$ measurements are available for many early-type
galaxies, these do not uniquely constrain the age of their stellar
population. A more powerful constraint is provided by the {\it
relative} rate at which the $M/L$ varies with time, or alternatively,
redshift: $[d(M/L) / (M/L)] / dz \propto d \log (M/L) / dz$. Because this is
a relative measure, it depends less strongly on uncertainties in the
IMF for low-mass stars. Of course, a dependence does remain on the
exact shape of the IMF for the higher-mass stars that produce most of
the light. Nonetheless, $d \log (M/L) / dz$ can be used to meaningfully
constrain the formation redshift of early-type galaxies (van Dokkum
\etal 1998).

Studies of $d \log (M/L) / dz$ have so far relied primarily on
measurements of the Fundamental Plane (FP) for galaxies in the nearby
and distant Universe (van Dokkum \& Franx 1996). The FP is a tight
planar relation between the global properties of early-type galaxies
in any three-dimensional parameter space spanned by quantities that
measure the characteristic galaxy size, velocity dispersion, and
surface brightness (Djorgovski \& Davis 1987; Dressler \etal
1987). For example, one might take the effective radius $r_{\rm eff}$,
the average velocity dispersion $\sigma_{\rm eff}$ inside the
effective radius, and the average surface brightness $I_{\rm eff}$
inside the effective radius. The existence of the FP can be understood
as a combination of the virial theorem and a power-law dependence of
$M/L$ on global galaxy properties (Dressler \etal 1987; Cappellari
\etal 2006a). A decrease in $M/L$ with redshift due to stellar 
population effects corresponds to an increase in $I_{\rm eff}$, and
consequently, a measurable decrease in the zeropoint of the FP.

Studies of FP evolution have provided important new insights into the
formation and evolution of early-type galaxies (see, e.g., the
following recent papers and references therein: Wuyts \etal 2004; Woo
\etal 2004; Moran \etal 2005; Treu \etal 2005a,b; van der Wel \etal
2004, 2005; di Serego Alighieri \etal 2005; Jorgensen \etal 2006). In van
Dokkum \& van der Marel (2006, hereafter vDvdM06), we presented
spectroscopy with the Low Resolution Imager and Spectrograph (LRIS) on
Keck of some two dozen galaxies in the intermediate-redshift ($z
\approx 0.5$) clusters CL3C295, CL0016+16 and CL1601+42. The sample
galaxies were selected to be bright enough for spectroscopy, and
visually classified from the Hubble Space Telescope (HST) images of
Dressler \etal (1997) and Smail \etal (1997) as early-type (and in
most cases elliptical) galaxies. We measured the integrated velocity
dispersions of the galaxies and analyzed existing HST images to infer
their characteristic photometric properties. This allowed a study of
the FP evolution of the three sample clusters. We combined our new
results with existing FP data for eleven additional clusters in the
redshift range $0.18 \leq z \leq 1.28$ as well as samples of field
early-type galaxies in the redshift range $0.32 \leq z \leq
1.14$. This provides the largest homogenized analysis of FP evolution
to date and implies a luminosity-weighted mean star formation redshift
$z_{*} = 2.01_{-0.17}^{+0.22}$ for massive ($M > 10^{11} \Msun$)
early-type galaxies in clusters. Field early-type galaxies in the same
mass range are only $4.1 \pm 2.0$\% younger, with $z_{*} =
1.95_{-0.08}^{+0.10}$. (These results assume that the IMF has a
``standard'' form and that the progenitor bias described by van Dokkum
\& Franx (2001) does not depend on environment; see vDvdM06 for
details).

The success and popularity of FP studies can be attributed at least in
part to the relative ease with which such studies can be
performed. Only global galaxy properties need to be measured, and
modeling of the internal structure of the sample galaxies is not
required. However, there are important caveats. In particular, values
of $d \log (M/L) / dz$ determined from FP evolution are correct only
if many simplifying assumptions are satisfied. Any potential evolution
of global galaxy properties other than $M/L$ (i.e., $\sigma_{\rm
eff}$, $r_{\rm eff}$ and $\Sigma_{\rm eff} \equiv (M/L)\times I_{\rm
eff} (r)$, the latter quantity being the average projected surface
mass density inside $r_{\rm eff}$) must either be absent or be such so
as to not affect the inferred $M/L$ evolution. Also, galaxies must
change homologously with redshift, if at all; they must not evolve in
quantities such as the shape of their three-dimensional contours, the
shape of their density profile with radius, or the shape of their
intrinsic dynamical structure. The possibility that evolution in any
of these quantities may in fact occur has left the interpretation of
FP studies somewhat uncertain. This possibility has come into sharp
focus through the fact that the inferred $M/L$ evolution of massive
early-type galaxies is well fit by models of passive evolution
following formation at high redshift, and by the fact that there
appears to be only a small age difference between massive early-type
galaxies in clusters and the field (vDvdM06). Although models can be
constructed that reproduce these results (e.g., Nagamine \etal 2005;
De Lucia \etal 2006), they do appear somewhat counter-intuitive given
the paradigm of hierarchical structure formation in the
Universe. Moreover, there exist sophisticated semi-analytical models
in which the evolution of galaxy $M/L$ and FP zeropoint are actually
quite different (Almeida, Baugh \& Lacey 2006). So there are many
reasons to critically question whether FP evolution does in fact
uniquely trace $M/L$ evolution.

The most direct way to address this question is to determine $M/L$
values for individual distant galaxies in the same way as has been
done locally, without resorting to FP assumptions.  This requires
construction of detailed dynamical models for high-quality spatially
resolved photometric and kinematic data. With these goals in mind we
obtained spectroscopic data in vDvdM06 that was deep enough to extract
spatially resolved rotation curves and velocity dispersion profiles
for the sample galaxies in CL3C295, CL0016+16 and CL1601+42. In van
der Marel \& van Dokkum (2006, hereafter Paper~I) we presented the
kinematical profiles and constructed detailed dynamical models to
interpret them. The models are axisymmetric and are based on solving
the Jeans equations of hydrostatic equilibrium under the assumption of
a two-integral distribution function $f=f(E,L_z)$, where $E$ is the
energy and $L_z$ the angular momentum around the symmetry
axis. Fitting of the models to the available HST imaging and observed
kinematical profiles yields two quantities: a normalized measure $k$
of the galaxy's rotation rate and the $M/L$ of the stellar population
(in rest-frame $B$ band solar units). The inferred values for these
quantities and their formal uncertainties were presented in Paper~I.
The implications of the inferred rotation rates were also discussed in
that paper. In the present paper we interpret the inferred $M/L$
values. To do this, we first compile and homogenize a comparison
sample of galaxies in the local Universe with reliable $M/L$
determinations from the literature. We then compare the $M/L$ values
for the intermediate-redshift cluster galaxies to those for the local
comparison sample to obtain a direct measure of the $M/L$ evolution of
elliptical galaxies.


\newcommand{\tablecontentssyserrors}{ 
\tablecaption{Sources of systematic uncertainty that affect
$\log(M/L)$ measurements\label{t:syserrors}}
\tablehead{
\colhead{ID} & \colhead{$\Delta \log(M/L)$} & 
\colhead{source of systematic uncertainty} & \colhead{Section} \\
}
\startdata
I   & $\pm 0.003$ & effect of uncertainties in $\Omega_{\rm m}$ and 
$\Omega_{\Lambda}$ at $z \approx 0.5$ & \ref{s:intro} \\
II  & $\pm 0.026$ & accuracy with which the SBF and Cepheid distance
scales have been aligned & \ref{ss:dist} \\ 
III & $\pm 0.037$ & accuracy of the Cepheid distance scale for local galaxies &
\ref{ss:dist} \\
IV  & $\pm 0.009$ & cosmic variance in $H_0$ on the scale $z \lta
0.1$ & \ref{ss:dist} \\
V   & $\pm 0.023$ & random uncertainty in $H_0$ due to finite sample sizes & 
\ref{ss:dist} \\
VI  & $\pm 0.020$ & difference in results from different dynamical modeling 
approaches & \ref{ss:MLerrors} \\
\enddata}

\newcommand{\tablecommsyserrors}{Column~(1) lists the roman numeral 
with which a particular systematic uncertainty is referred to in the
text. Column~(2) lists the size of the systematic
uncertainty. Column~(3) lists the source of the systematic
uncertainty. Column~(4) lists the number of the section in which the
uncertainty is discussed.}

\ifemulate
\begin{deluxetable*}{lclc}
\tabletypesize{\footnotesize}
\tablecontentssyserrors
\tablecomments{\tablecommsyserrors}
\end{deluxetable*}
\fi


The layout of this paper is as follows. Section~\ref{s:local}
discusses the compilation of the local comparison sample of galaxies
with dynamically inferred $M/L$ values. Section~\ref{s:MLevolution}
compares the $M/L$ values for the galaxies in the
intermediate-redshift sample from Paper~I to those from the local
compilation. This yields $d \log (M/L) / dz$, which is compared to the
FP evolution of the same sample derived in
vDvdM06. Section~\ref{s:slopeevol} discusses how the $M/L$ evolution
depends on galaxy dispersion (or similarly mass), both in the present
study and in the FP analysis. The uncertainties in both methods are
discussed, as well as the implications for the accuracies of the
inferred $M/L$ evolution. Section~\ref{s:conc} presents a summary and
discussion of the results.

The dynamically inferred $M/L$ of a galaxy is inversely proportional
to the assumed distance. Throughout this paper (as in Paper~I) we
assume a cosmology with $\Omega_{\rm m} = 0.27$, $\Omega_{\Lambda} =
0.73$ (the values obtained by the Wilkinson Microwave Anisotropy
Probe; Spergel \etal 2003) and $H_0 = 71 \kms \Mpc^{-1}$ (the value
obtained by the HST Cepheid Key Project; Section~7 of Freedman \etal
2001). The uncertainties in $\Omega_{\rm m}$, $\Omega_{\Lambda}$ can
be estimated to be $\sim 0.02$ (Spergel \etal 2003). At $z \approx
0.5$ this introduces an uncertainty of only $\Delta_{\rm I} = \pm
0.003$ in $\log(M/L)$. (Here and henceforth we denote sources of
systematic uncertainty in $\log(M/L)$ with a roman numeral
subscript. A summary listing of all sources of systematic uncertainty
encountered in this paper is presented in Table~\ref{t:syserrors}.)
The uncertainties in $H_0$ and their effect on $\log(M/L)$ will be
addressed later.

\section{Local Galaxy Mass-to-Light Ratio Comparison Sample}
\label{s:local}

\subsection{Dynamical Modeling Sources}
\label{ss:local}

To study the $M/L$ evolution with redshift we need a comparison sample
of dynamically inferred $M/L$ values for {\it nearby} early-type
galaxies. We restrict our attention here to five detailed dynamical
modeling studies that addressed relatively large samples: van der
Marel (1991, hereafter vdM91); Magorrian \etal (1998, hereafter M98);
Kronawitter \etal (2000, hereafter K00); Gebhardt \etal (2003,
hereafter G03); and Cappellari \etal (2006a, hereafter C06). These
studies differ from each other in many ways, both in terms of the
quality and nature of the data that were used, and in the methods and
sophistication of the modeling. In particular:

\begin{itemize}

\item vdM91 and K00 used ground-based photometry, whereas M98,
G03 and C06 used a combination of both HST and ground-based
photometry;

\item vdM91, M98, K00 and C06 used ground-based spectroscopy, whereas
G03 used a combination of ground-based and HST spectroscopy;

\item vdM91, M98, K00 and G03 used long-slit spectroscopy along one or
more slit position angles, whereas C06 used fully two-dimensional
integral-field spectroscopy;

\item K03 constructed spherical dynamical models (they restricted their 
sample to galaxies that are almost circular in projection on the sky),
whereas vdM91, M98, G03 and C06 constructed axisymmetric models;

\item vdM91 and M98 constructed two-integral models using the Jeans
equations, K00 constructed models using an expansion around a set of
known basis distribution functions, and G03 and C06 constructed fully
general models using numerical orbit superposition;

\item K00 included kinematical data in the central few arcsec in their
fits, but did not allow for the possible gravitational contribution of
a central BH; vdM91 also did not include a BH in the models but
excluded the central few arcsec from the fit; C06 included a BH of
fixed mass in the models but still excluded the central few arcsec
from the fit; and M98 and G03 included BHs in their models and fitted
their masses by using the data in the central few arcsec;

\item vdM91, M98, G03 and C06 assumed a constant value of $M/L$ with
radius, whereas K03 explicitly included (and optimized) the
contribution of a dark halo;

\item K03 determined $M/L$ values in the $B$-band, M98 and G03 in the
$V$-band,\footnote{G03 reported the $M/L$ for one galaxy, NGC 4564, in
the $I$-band.} vdM91 in the Johnson $R_{\rm J}$ band, and C06 in the
$I$-band.

\end{itemize}

We purposely included the results from all five studies, rather than
to retain merely a smaller sample composed of the most recent or most
accurate results. One advantage of this it that by comparing the
results from the different studies we are able to put firm limits on
various kinds of potential systematic uncertainties.

\subsection{Distances}
\label{ss:dist}

It is important for our study to use a set of galaxy distances that is
as accurate and homogeneous as possible. For this we follow the
example of G03 and C06 by using the compilation of Tonry \etal (2001,
hereafter T01). They obtained distances to 300 nearby galaxies, mostly
of early type, using the surface brightness fluctuation (SBF) method.
We removed from our initial sample of 81 total galaxies in vdM91, M98,
K00, G03 and C06 the 17 galaxies that are not part of the T01
sample. We rejected two more galaxies (M31 and NGC 4594) because they
are not early-type galaxies (which we define here as Hubble type $T <
0$) and two other galaxies (NGC 3384 and NGC 7332) for the reasons
discussed in Section~\ref{ss:dispersions}. Table~\ref{t:local} lists
the final sample of 60 galaxies and includes for each galaxy, among
other things, the Hubble type $T$ and the adopted distance modulus and
its uncertainty. The average distance for the sample galaxies is $21.6
\Mpc$. This corresponds to $\langle z \rangle = 0.005$, given the Hubble 
constant listed in Section~\ref{s:intro}.

There are 6 spiral galaxies with bulges in common between the SBF
sample of T01 and the sample of galaxies for which Cepheid distances
are available from the HST Cepheid Key Project. T01 calibrated the
zeropoint of their SBF method so as to set to zero the {\it median}
SBF vs.~Cepheid distance modulus residual for these 6 galaxies (see
Appendix B of Tonry \etal 2000). This was done using Cepheid results
available in 2000. These Cepheid distances have since been improved
(Freedman \etal 2001). Application of the same calibration methodology
to the new Cepheid distances implies an SBF zeropoint that is larger
by $+0.06$ mag (e.g., Mei \etal 2005). To account for this, we added a
correction $\Delta_{\rm T01} = -0.06$ mag to all the distance moduli
in T01 (i.e., moving the galaxies closer).

For the purposes of our study it is important to understand the
accuracy with which the SBF and Cepheid distance scales have been
aligned. If one uses the weighted average rather than the median
statistic to align the 6 spiral galaxies, then the SBF zeropoint
changes by $-0.12$. Alternatively, if one chooses not to align the
distance scales using individual galaxies at all, but instead using
distances to groups of galaxies (Ferrarese \etal 2000), then this
yields a change of $-0.13$ to the SBF zeropoint (Tonry \etal
2000). And finally, if one changes the SBF zeropoint by $+0.10$, then
better agreement is obtained with predictions of population synthesis
models (Jensen \etal 2003).\footnote{This zeropoint also happens to
align the SBF distance scale with the Cepheid distance scale {\it if}
one assumes that the Cepheid properties have no metallicity
dependence. But that is not what Freedman \etal (2001) assumed in
their final calibration. Comparisons of TRGB and Cepheid distances
also indicate a Cepheid metallicity dependence (Sakai \etal 2004).
The shift advocated by Jensen \etal (2003) was used by C06 in their
study of $M/L$ values of nearby galaxies.} Based on these
considerations we assume that the systematic uncertainty on
$\Delta_{\rm T01}$ is $\pm 0.13$ mag. This corresponds to an
uncertainty $\Delta_{\rm II} = \pm 0.026$ in $\log(M/L)$.

After its calibration to agree with Cepheids, the SBF distance scale
is subject to all the same absolute distance scale uncertainties that
are inherent to the Cepheid distance scale. These are summarized in
Section~8 and Table~14 of Freedman \etal (2001). Most of the
uncertainties affect both the distances of local galaxies, as well as
the inferred value of $H_0$. They include: the distance to the LMC,
the photometric zeropoint of the HST/WFPC2 used for the Cepheid
studies, the accuracy of the reddening and metallicity corrections
applied to the Cepheids, and biases introduced by crowding and the
magnitude limit of the sample. When added in quadrature, as in
Freedman \etal (2001), these effects introduce a 9\% systematic
uncertainty in the galaxy distances. This corresponds to a difference
$\Delta_{\rm III} = \pm 0.037$ in $\log(M/L)$. This uncertainty must be
taken into account, e.g., when comparing the $M/L$ values of local
galaxies to the predictions of stellar population synthesis models.

The systematic Cepheid distance scale uncertainties listed in the
previous paragraph do {\it not} affect a relative comparison of $M/L$
values of nearby and distant galaxies. Any shift in the overall
distance scale of the Universe would affect both types of galaxies
equally, and would therefore cancel out when evaluating a difference
in $\log(M/L)$. The only distance scale uncertainties that do affect
such a comparison are those that impact the estimate of $H_0$, but not
the Cepheid distances to local galaxies. These uncertainties have both
a systematic and a random component.  The systematic component is due
to cosmic variance, i.e., the effect that local samples of galaxies
may yield an estimate of $H_0$ that differs from the true cosmic value as
a result of bulk flows. Based on the discussion in Section~8.6 of
Freedman \etal (2001) we estimate this systematic component to be less
than 2\% on the scales $z \lta 0.1$ over which Type Ia supernovae have
been used to calibrate $H_0$. This corresponds to a difference
$\Delta_{\rm IV} = \pm 0.009$ in $\log(M/L)$. Random uncertainties in
$H_0$, as opposed to systematic ones, result from finite sample sizes
and scatter between results from individual measurements. Based on
Section~7 of Freedman \etal (2001) we estimate the random uncertainty
to be $\Delta H_0 = \pm 4 \kms \Mpc^{-1}$. This corresponds to
$\Delta_{\rm V} = \pm 0.023$ in $\log(M/L)$.


\newcommand{\tablecontentslocal}{
\tablecaption{Local Galaxy Sample\label{t:local}}
\tablehead{
\colhead{Galaxy} & \colhead{Type} & \colhead{$m-M$} & 
\colhead{$\log \sigma_{\rm eff}$} & 
\colhead{$\log M/L_B$} & 
\colhead{vdM91} & \colhead{M98} & \colhead{K00} & 
\colhead{G03} & \colhead{C06} \\
\colhead{(1)} & \colhead{(2)} & \colhead{(3)} & \colhead{(4)} &
\colhead{(5)} & \colhead{(6)} & \colhead{(7)} & \colhead{(8)} &
\colhead{(9)} & \colhead{(10)} \\
}
\startdata
NGC 0221  & -6  & $24.49 \pm 0.08$ & -0.481  & $0.403 \pm 0.053$ &     &  X  &     &     &  X \\
NGC 0524  & -1  & $31.84 \pm 0.20$ &  0.064  & $1.022 \pm 0.067$ &     &     &     &     &  X \\
NGC 0636  & -5  & $32.31 \pm 0.16$ & -0.134  & $0.694 \pm 0.116$ &  X  &     &     &     &    \\
NGC 0720  & -5  & $32.15 \pm 0.17$ &  0.049  & $1.034 \pm 0.116$ &  X  &     &     &     &    \\
NGC 0821  & -5  & $31.85 \pm 0.17$ & -0.022  & $0.881 \pm 0.049$ &     &  X  &     &  X  &  X \\
NGC 1052  & -5  & $31.38 \pm 0.27$ & -0.020  & $1.026 \pm 0.116$ &  X  &     &     &     &    \\
NGC 1379  & -5  & $31.45 \pm 0.15$ & -0.212  & $0.582 \pm 0.116$ &  X  &     &     &     &    \\
NGC 1395  & -5  & $31.85 \pm 0.16$ &  0.069  & $0.921 \pm 0.116$ &  X  &     &     &     &    \\
NGC 1399  & -5  & $31.44 \pm 0.16$ &  0.155  & $1.024 \pm 0.059$ &  X  &  X  &  X  &     &    \\
NGC 1404  & -5  & $31.55 \pm 0.19$ &  0.026  & $0.896 \pm 0.116$ &  X  &     &     &     &    \\
NGC 1407  & -5  & $32.24 \pm 0.26$ &  0.097  & $1.057 \pm 0.116$ &  X  &     &     &     &    \\
NGC 1439  & -5  & $32.07 \pm 0.15$ & -0.149  & $0.777 \pm 0.116$ &  X  &     &     &     &    \\
NGC 1549  & -5  & $31.41 \pm 0.18$ & -0.027  & $0.763 \pm 0.116$ &  X  &     &     &     &    \\
NGC 1700  & -5  & $33.17 \pm 0.16$ &  0.039  & $0.842 \pm 0.116$ &  X  &     &     &     &    \\
NGC 2434  & -5  & $31.61 \pm 0.29$ & -0.026  & $0.938 \pm 0.117$ &     &     &  X  &     &    \\
NGC 2778  & -5  & $31.74 \pm 0.30$ & -0.097  & $0.975 \pm 0.071$ &     &  X  &     &  X  &    \\
NGC 2974  & -5  & $31.60 \pm 0.24$ &  0.074  & $0.989 \pm 0.067$ &     &     &     &     &  X \\
NGC 3115  & -3  & $29.87 \pm 0.09$ &  0.109  & $0.995 \pm 0.085$ &     &  X  &     &     &    \\
NGC 3156  & -2  & $31.69 \pm 0.14$ & -0.420  & $0.349 \pm 0.067$ &     &     &     &     &  X \\
NGC 3193  & -5  & $32.60 \pm 0.18$ & -0.025  & $0.711 \pm 0.117$ &     &     &  X  &     &    \\
NGC 3377  & -5  & $30.19 \pm 0.09$ & -0.194  & $0.571 \pm 0.049$ &     &  X  &     &  X  &  X \\
NGC 3379  & -5  & $30.06 \pm 0.11$ & -0.009  & $0.821 \pm 0.044$ &  X  &  X  &  X  &     &  X \\
NGC 3414  & -2  & $31.95 \pm 0.33$ &  0.002  & $0.925 \pm 0.067$ &     &     &     &     &  X \\
NGC 3557  & -5  & $33.24 \pm 0.22$ &  0.112  & $0.842 \pm 0.116$ &  X  &     &     &     &    \\
NGC 3608  & -5  & $31.74 \pm 0.14$ & -0.058  & $0.851 \pm 0.045$ &  X  &  X  &     &  X  &  X \\
NGC 3640  & -5  & $32.10 \pm 0.13$ & -0.092  & $0.694 \pm 0.117$ &     &     &  X  &     &    \\
NGC 4150  & -2  & $30.63 \pm 0.24$ & -0.438  & $0.378 \pm 0.067$ &     &     &     &     &  X \\
NGC 4168  & -5  & $32.39 \pm 0.42$ & -0.089  & $0.929 \pm 0.069$ &     &  X  &  X  &     &    \\
NGC 4261  & -5  & $32.44 \pm 0.19$ &  0.123  & $1.146 \pm 0.116$ &  X  &     &     &     &    \\
NGC 4278  & -5  & $30.97 \pm 0.20$ &  0.064  & $0.996 \pm 0.048$ &     &  X  &  X  &     &  X \\
NGC 4291  & -5  & $32.03 \pm 0.32$ &  0.095  & $0.943 \pm 0.071$ &     &  X  &     &  X  &    \\
NGC 4374  & -5  & $31.26 \pm 0.11$ &  0.151  & $0.992 \pm 0.052$ &  X  &     &  X  &     &  X \\
NGC 4406  & -5  & $31.11 \pm 0.14$ &  0.047  & $0.904 \pm 0.116$ &  X  &     &     &     &    \\
NGC 4458  & -5  & $31.12 \pm 0.12$ & -0.403  & $0.638 \pm 0.067$ &     &     &     &     &  X \\
NGC 4459  & -1  & $30.98 \pm 0.22$ & -0.089  & $0.710 \pm 0.067$ &     &     &     &     &  X \\
NGC 4472  & -5  & $31.00 \pm 0.10$ &  0.105  & $1.029 \pm 0.059$ &  X  &  X  &  X  &     &    \\
NGC 4473  & -5  & $30.92 \pm 0.13$ & -0.018  & $0.816 \pm 0.049$ &     &  X  &     &  X  &  X \\
NGC 4486  & -4  & $30.97 \pm 0.16$ &  0.208  & $1.139 \pm 0.044$ &  X  &  X  &  X  &     &  X \\
NGC 4494  & -5  & $31.10 \pm 0.11$ & -0.239  & $0.719 \pm 0.082$ &  X  &     &  X  &     &    \\
NGC 4526  & -2  & $31.08 \pm 0.20$ &  0.051  & $0.836 \pm 0.067$ &     &     &     &     &  X \\
NGC 4550  & -2  & $30.94 \pm 0.20$ & -0.252  & $0.685 \pm 0.067$ &     &     &     &     &  X \\
NGC 4552  & -5  & $30.87 \pm 0.14$ &  0.099  & $0.986 \pm 0.053$ &     &  X  &     &     &  X \\
NGC 4564  & -5  & $30.82 \pm 0.17$ & -0.129  & $0.792 \pm 0.071$ &     &  X  &     &  X  &    \\
NGC 4589  & -5  & $31.65 \pm 0.22$ & -0.006  & $1.071 \pm 0.117$ &     &     &  X  &     &    \\
NGC 4621  & -5  & $31.25 \pm 0.20$ &  0.025  & $0.846 \pm 0.053$ &     &  X  &     &     &  X \\
NGC 4636  & -5  & $30.77 \pm 0.13$ & -0.069  & $1.089 \pm 0.059$ &  X  &  X  &  X  &     &    \\
NGC 4649  & -5  & $31.07 \pm 0.15$ &  0.188  & $1.053 \pm 0.060$ &  X  &  X  &     &  X  &    \\
NGC 4660  & -5  & $30.48 \pm 0.19$ & -0.034  & $0.873 \pm 0.053$ &     &  X  &     &     &  X \\
NGC 4697  & -5  & $30.29 \pm 0.14$ & -0.120  & $0.918 \pm 0.086$ &  X  &     &     &  X  &    \\
NGC 5813  & -5  & $32.48 \pm 0.18$ &  0.061  & $0.990 \pm 0.067$ &     &     &     &     &  X \\
NGC 5845  & -5  & $32.01 \pm 0.21$ &  0.077  & $0.872 \pm 0.060$ &     &     &     &  X  &  X \\
NGC 5846  & -5  & $31.92 \pm 0.20$ &  0.104  & $1.115 \pm 0.052$ &  X  &     &  X  &     &  X \\
NGC 6703  & -3  & $32.07 \pm 0.29$ & -0.097  & $0.749 \pm 0.117$ &     &     &  X  &     &    \\
NGC 7144  & -5  & $31.89 \pm 0.12$ & -0.074  & $0.840 \pm 0.116$ &  X  &     &     &     &    \\
NGC 7145  & -5  & $31.79 \pm 0.21$ & -0.218  & $0.761 \pm 0.082$ &  X  &     &  X  &     &    \\
NGC 7192  & -4  & $32.83 \pm 0.32$ & -0.080  & $0.665 \pm 0.117$ &     &     &  X  &     &    \\
IC 1459   & -5  & $32.27 \pm 0.28$ &  0.145  & $0.912 \pm 0.116$ &  X  &     &     &     &    \\
NGC 7457  & -3  & $30.55 \pm 0.21$ & -0.444  & $0.550 \pm 0.060$ &     &     &     &  X  &  X \\
NGC 7507  & -5  & $31.93 \pm 0.17$ &  0.042  & $0.801 \pm 0.082$ &  X  &     &  X  &     &    \\
NGC 7619  & -5  & $33.56 \pm 0.31$ &  0.175  & $1.002 \pm 0.116$ &  X  &     &     &     &    \\
\enddata}

\newcommand{\tablecommlocal}{Column~(1) lists the galaxy name. 
Columns~(2) and~(3) list the morphological type and distance modulus
from T01; the T01 values were shifted by $-0.06$ mag to align the SBF
distance scale with the Cepheid distance scale. Column~(4) lists the
base-10 logarithm of $\sigma_{\rm eff}$ in km/s, either taken directly
from C06 or otherwise calculated from the data in F89. We assume the
uncertainties on $\log (\sigma_{\rm eff})$ to be $0.021$. Column~(5)
lists the base-10 logarithm of $(M/L_B)$ and its uncertainty, obtained
through weighted combination of the results from different
studies. (The uncertainties should not be used for studies of the
intrinsic scatter in relations between $M/L$ and other global galaxy
properties, because they were derived under the assumption that the
intrinsic scatter in the correlation with $\sigma_{\rm eff}$ is
negligible.)  Columns (6)--(10) indicate the original dynamical
modeling sources from which the $M/L$ values were obtained: van der
Marel (1991, vdM91); Magorrian \etal (1998, M98); Kronawitter \etal
(2000, K00); Gebhardt \etal (2003, G03); or Cappellari \etal (2006a,
C06).}

\ifemulate
\begin{deluxetable*}{lccrcccccc}
\tabletypesize{\small}
\tablecontentslocal
\tablecomments{\tablecommlocal}
\end{deluxetable*}
\fi


\subsection{Transformations to the $B$-band}
\label{ss:colors}

For comparison to results at intermediate redshifts it is important to
have access to $M/L$ values in the rest-frame $B$-band. We therefore
took the published values\footnote{Some studies list multiple values
of the $M/L$ obtained under slightly different assumptions. For vdM91
we take the values labeled $\Upsilon_R^{\rm imp}$ in his Table 2. For
K00 we take the central value labeled ``best $M/L_B^c$'' in their
Table 7. For C06 we take the values from their Schwarzschild models,
labeled $(M/L)_{\rm Schw}$ in their Table 1, as opposed to the values
from their two-integral models.} and transformed them to $B$-band
mass-to-light ratios $M/L_B$ for the distances $D$ discussed in
Section~\ref{ss:dist} using
\begin{equation}
\label{MLtrans}
   M/L_B = (M/L_F) \> (D_{\rm orig}/D) \> 10^{0.4 [(B-F) - (B-F)_{\rm
           \odot}]} .
\end{equation}
Here $M/L_F$ is the mass-to-light ratio in some other band $F$ for the
distance $D_{\rm orig}$, as given in the original literature source;
$B-F$ is the relevant broad-band color of the galaxy, and $(B-F)_{\rm
\odot}$ is the color of the sun.

For the transformation in equation~(\ref{MLtrans}) we needed colors of
the sample galaxies in various bands. For $B-V$ we used the data of
Faber \etal (1989, hereafter F89), who presented data for almost 600
nearby early-type galaxies. Their colors refer to apertures $\leq
30''$ in diameter around the galaxy center. The average color for the
galaxies in our sample is $\langle B-V \rangle = 0.95$. We used this
average color for the 8 galaxies in our final sample for which a $B-V$
was needed but for which none was available from F89. For those
galaxies for which we needed the color $B-R_{\rm J}$ we used the
transformation $B-R_{\rm J} = 1.839(B-V) + 0.064$. This was obtained
from the equations $B-R_{\rm J} = 1.14 (B-R_{\rm C}) + 0.04$ and
$V-R_{\rm C} = 0.613(B-V) + 0.021$ derived for early-type galaxies by
Peletier \etal (1990). The subscripts in these equations refer to the
Johnson and the Cousins $R$-band, respectively. The implied average
color of the sample is $\langle B-R_{\rm J} \rangle = 1.81$.

For the transformation of $M/L$ values in the $I$-band we needed the
$B-I$ colors of the galaxies. We calculated those as the sum of $B-V$
and $V-I$, with the former from F89 and the latter from T01. However,
a small correction was needed to the T01 colors because they apply to
an annular region around the galaxy center.  The dynamical $M/L$
determinations are therefore more heavily weighted towards the galaxy
center than the T01 colors. The central region that was excluded by
T01 has an average diameter of $24''$ for those galaxies in our sample
for which the original $M/L$ was given in the $I$-band. The average
color of these galaxies is $\langle (V-I)_{\rm T01} \rangle =
1.16$. Early-type galaxies become redder towards their centers, and
this implies that the T01 colors are bluer than the ones that should
be used in equation~(\ref{MLtrans}). As a simple correction for this
we used $V-I = (V-I)_{\rm T01} + \epsilon$. We applied this equation
on a galaxy-by-galaxy basis, but the small constant $\epsilon$ was
chosen to be a fixed number. To set its value we compared to the
predictions of stellar population synthesis models.  The central
colors $\langle B-V \rangle = 0.95$ and $\langle B-R_{\rm J} \rangle =
1.81$ of early-type galaxies in our sample are fit to within $0.01$
mag by a $10^{9.97}$ yr old single age stellar population with a
Chabrier IMF of solar metallicity and solar abundances ratios (Bruzual
\& Charlot 2003).\footnote{This does not necessarily mean that the
stellar populations are indeed on average $10^{9.97}$ yr old, and have
solar metallicity and solar abundances ratios. But it does mean that
such models predict the correct continuum slope when compared to
observations of the centers of nearby early-type galaxies. The models
can therefore be used to estimate the broad-band colors in filter
combinations for which observations are not directly available.}  Such
a population has $V-I = 1.20$ (Bruzual \& Charlot 2003). To make the
average T01 colors consistent with these stellar population
predictions we therefore choose $\epsilon = 0.04$. The sign and size
of this offset is consistent with our understanding of the $V-I$ color
gradients in the centers of early-type galaxies (Lauer
\etal 2005).

In equation~(\ref{MLtrans}) we also needed the colors of the
sun. Throughout this paper we use the solar absolute magnitudes
compiled by Binney \& Merrifield (1998): $M_{\odot} = 5.48 (B), 4.83
(V), 4.42 (R_{\rm C})$, and $4.08 (I)$. For the transformation of the
vdM91 results we needed also the solar absolute magnitude in the
Johnson $R$-band. For this we used the color transformation $R_{\rm J}
= R_{\rm C} - 0.12 (B-R_{\rm C}) - 0.07$ (Davis \etal 1985) to obtain
$M_{\odot} = 4.22 (R_{\rm J})$.\footnote{vdM91 had used $M_{\odot} =
4.31 (R_{\rm J})$, so we corrected his results to $M_{\odot} = 4.22
(R_{\rm J})$ before application of equation~(\ref{MLtrans}).}

\subsection{Velocity Dispersions}
\label{ss:dispersions}

For the analysis and interpretation of the results it is important to
have access also to other characteristic quantities for the sample
galaxies. The velocity dispersion is particularly useful, because it
has been found to correlate strongly with the galaxy $M/L$ (C06). To
characterize the dispersion we chose the value $\sigma_{\rm eff}$
inside an aperture of size equal to the effective radius $r_{\rm
eff}$. C06 directly measured this quantity for the galaxies in their
sample from their own data, and we adopted their values for those
galaxies. For the remainder of the galaxies in our sample we started
from the values $\sigma_{\rm nuc}$ given by F89. These are averages of
observed values from different observational setups, corrected to the
size of a few-arcsec aperture at the distance of Coma. From these
values we estimated the dispersion in an aperture of size $r_{\rm
eff}$, using the values of $r_{\rm eff}$ also given by F89 and the
correction formulae given in Jorgensen \etal (1995b). The two galaxies
NGC 3384 and NGC 7332 did not have data available from either C06 or
F89, and we removed these galaxies from the sample.

The uncertainties in $\sigma_{\rm eff}$ are probably not dominated by
random uncertainties due to the finite $S/N$ of the spectroscopy, but
by systematic uncertainties associated with template mismatch,
continuum subtraction, and other issues. For the high-quality integral
field data of C06 we followed those authors and estimated the
uncertainties in $\sigma_{\rm eff}$, somewhat conservatively, to be
$\sim 5$\% (see also Tremaine \etal 2002). This implies $\Delta \log
\sigma_{\rm eff} = 0.021$. For those galaxies for which a dispersion
estimate is available also from the F89 data we find that the
residuals $\delta \log \sigma_{\rm eff}
\equiv \log \sigma_{\rm eff,F89} - \log \sigma_{\rm eff,C06}$ are on 
average consistent with zero: $\langle \delta \log \sigma_{\rm eff}
\rangle = -0.007 \pm 0.008$. The RMS scatter in the residuals is
$0.031$.\footnote{This scatter was obtained by comparing only galaxies
with $\sigma_{\rm eff} > 100 \kms$, as appropriate for all the
galaxies in our sample for which we actually use the F89 data.  We did
find larger residuals of $\sim 0.1$ dex for the few galaxies with
$\sigma_{\rm eff} < 100 \kms$.} This is approximately what would be
expected if the $\sigma_{\rm eff}$ values inferred from the F89 data
also have uncertainties of $\sim 5$\%, which is therefore what we assumed.


\newcommand{\figcaplocal}{Correlation of $M/L$ (in solar units)
with $\sigma_{\rm eff}$ (in km/s) for nearby early-type galaxies. The
$M/L$ measurements from the dynamical studies of van der Marel (1991,
vdM91), Magorrian \etal (1998, M98), Kronawitter \etal (2000, K00),
Gebhardt \etal (2003, G03) and Cappellari \etal (2006a, C06) are shown
in separate panels.  All measurements were transformed to the $B$ band
and to a homogeneous set of SBF distances. The bottom-right panel
contains the combined dataset from Table~\ref{t:local}. The solid line
(the same in each panel) is the best fit to this combined data
set.\label{f:local}}

\ifemulate
\begin{figure*}
\epsfxsize=0.8\hsize
\centerline{\epsfbox{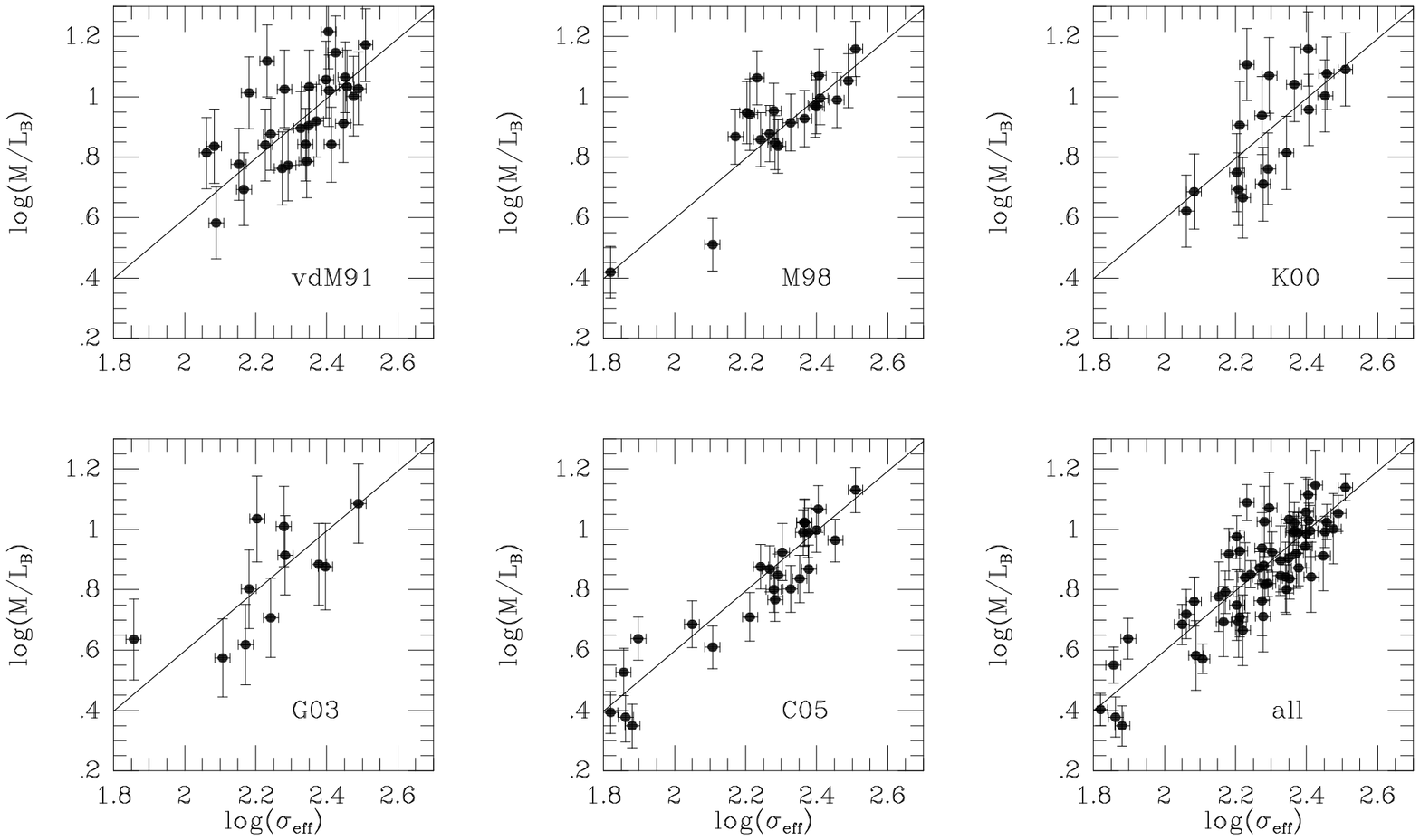}}
\figcaption{\figcaplocal}
\end{figure*}
\fi


Table~\ref{t:local} lists the inferred $\sigma_{\rm eff}$ for all
galaxies. The table does not list other galaxy parameters that we will
not use here. However, we note that various other quantities can be
obtained relatively easily. Effective radii are available from C06 or
F89. The latter authors also provide $I_{\rm eff}$, the average
$B$-band surface brightness inside the effective radius. The total
$B$-band luminosity can be estimated from $L = 2 \pi r_{\rm eff}^2
I_{\rm eff}$, or it can be obtained from the apparent $B$-band
magnitudes listed in, e.g., the RC3 (de Vaucouleurs \etal 1991). A
characteristic mass\footnote{The mass $M \equiv (M/L) \times L$ is
smaller than the total galaxy mass, because it doesn't include all the
contribution from a dark halo. However, it may contain some
contribution from dark matter (K00, C06), so it is probably larger
than the total mass in stars.} can be obtained upon multiplication by
the $M/L$ listed in the table.

\subsection{Mass-to-Light Ratio Accuracies and Correlations}
\label{ss:MLerrors}

The random uncertainties on the $M/L$ estimates can be divided broadly
in two components: distance uncertainties and dynamical
uncertainties. The random distance uncertainties come from the finite
accuracy of the SBF measurements. They contribute in quadrature
$\Delta \log (M/L) = 0.2 \Delta (m-M)$, where $\Delta (m-M)$ is the
uncertainty in the distance modulus from T01. The dynamical
uncertainties can come from a large variety of sources, e.g.,
shortcomings in the kinematical data or their spatial coverage, or
limitations in the modeling or its underlying assumptions. As a
result, they are generally poorly quantified. We quantify the random
dynamical uncertainties through a parameter $E$ which for simplicity
we assume to be a constant for all the measurements in a given
dynamical study. We then set the total random uncertainty $\Delta \log
(M/L)$ for a measurement in a given study equal to the sum of the
random distance uncertainty for the galaxy in question and the value
of $E$ for that study.

To estimate the random uncertainty $E$ for each study we use the fact that
the $M/L$ correlates with other global galaxy properties. Previous
work has shown that there are good correlations with either luminosity
or mass (vdM91, M98), as expected to explain the ``tilt'' of the
FP. More recently, C06 showed that the $M/L$ correlates even better,
i.e., with lower scatter, with $\sigma_{\rm eff}$. This is the
correlation that we will use here. For each individual dynamical study
we selected the $M/L_B$ values for the galaxies in our final sample
and then fitted a straight line of the form
\begin{equation}
\label{fitCapel}
  \log (M/L_B) = Z + S \log(\sigma_{\rm eff}/[200 \kms]) ;
\end{equation}
here ``Z'' is short for zeropoint, and ``S'' is short for slope.  We
will refer to this as the $M/L$--$\sigma$ relation. The fit was
performed with the routine fitexy of Press \etal (1992), which takes
into account the uncertainties in both independent variables. The
value of $E$ for each study was then chosen so as to yield a $\chi^2$
of the fit that is equal to the number of degrees of freedom. This
assumes implicitly that there is no intrinsic scatter in the
correlation, which is conservative in the sense that it yields the
largest possible random uncertainties. Application of this procedure
yields $E=0.116$ dex for vdM91, $E = 0.085$ dex for M98, $E=0.117$ dex
for K00, $E = 0.128$ for G03, and $E = 0.067$ dex for
C06. Figure~\ref{f:local} shows the final $B$-band $M/L$ values and
their random uncertainties as a function of $\sigma_{\rm eff}$ for all
five of the individual dynamical studies.

To obtain a combined $M/L_B$ estimate and uncertainty for each
individual galaxy we used a two-step procedure. First we took the
weighted average of the $M/L_B$ values inferred from the different
studies of that galaxy, using the uncertainties $E$ listed above to
set the weights. Then we increased the uncertainty in the weighted
average by adding in quadrature the random uncertainty $\Delta \log
(M/L) = 0.2 \Delta (m-M)$ introduced by distance
uncertainties. Table~\ref{t:local} lists the final $M/L_B$ estimates
thus obtained for all galaxies.

The bottom right panel of Figure~\ref{f:local} shows the best straight
line fit to the final combined data set. It has parameters $Z = 0.896
\pm 0.010$ and $S = 0.992 \pm 0.054$. This same line is
shown as a solid line in all panels of Figure~\ref{f:local}. The slope
of the relation differs significantly from the best-fit slope $S_I
[{\rm C06}] = 0.82 \pm 0.06$ found by C06 for the $I$-band. This is
not due to the use of a different galaxy sample. If we perform a
linear fit to only the $M/L_B$ values derived from the C06 study we
infer $S [{\rm C06}] = 0.991 \pm 0.076$, consistent with the value of $S$
inferred for the full sample. The difference in slope between the $I$-
and $B$-bands is therefore real, and is due to the well-known fact
that galaxies of low dispersion (or mass) tend to bluer than those of
high dispersion.

Figure~\ref{f:local} shows that the different studies are all entirely
consistent with each other. When the data from each study are fitted
individually with a straight line, the best fit slope is always
consistent with the value $S = 0.992 \pm 0.054$ inferred for the full
sample to within $1.2\sigma$ or better. When the data from each study
are fitted individually with a straight line of fixed slope $S =
0.992$, the inferred zeropoints are: $Z [{\rm vdM91}] = 0.914 \pm
0.023$, $Z [{\rm M98}] = 0.912 \pm 0.021$, $Z [{\rm K00}] = 0.897\pm
0.029$, $Z [{\rm G03}] = 0.894 \pm 0.040$, and $Z [{\rm C06}] = 0.876
\pm 0.017$. These zeropoints are all consistent with the value $Z = 0.896
\pm 0.010$ inferred for the full sample to within $1.2\sigma$ or better. 
The results from the C06 study have the smallest scatter in
Figure~\ref{f:local}, as quantified already by the parameter
$E$. Judged also from the sophistication and homogeneity of their
analysis, their results are probably the most reliable of the five
studies that we have included in our sample. On the other hand, the
color transformations that we had to apply to transform their $I$-band
results to the $B$-band are probably more uncertain than the
transformations that we had to apply to some of the other
studies. This may be the root cause of the fact that the zeropoint for
the C06 data is offset from that for the full sample by
$-0.020$. Either way, the more important conclusion in this context is
that the zeropoint of the relation is quite robust.\footnote{An
alternative way to assess differences in zeropoint between different
studies is to analyze the $M/L$ values for those galaxies that have
measurements from more than one study. We have performed such an
analysis and found zeropoint differences that are consistent with
those obtained from the $M/L$--$\sigma$ relation.}  The use of vastly
different data and models among 5 different studies does not alter the
zeropoint of the relation by more than $\Delta_{\rm VI} = \pm
0.020$. We adopt this as the systematic uncertainty in our knowledge
of the zeropoint $Z$ from dynamical modeling limitations. The
robustness of $M/L$ estimates from different modeling approaches is
consistent with several findings reported by C06. For example, they
find that the results of axisymmetric modeling generally do not depend
strongly on the assumed inclination. They also find that even though
two-integral and three-integral models yield subtly different $M/L$
estimates, there is little bias in the inferred average $M/L$ for a
sample that has a typical range of $M/L$ values.\footnote{The averages
of all the $\log M/L$ estimates in the C06 sample for two- and
three-integral models respectively agree to within $0.003$.}

We are now in a position to combine all the various sources of
systematic uncertainty that enter into the zeropoint $Z_B$ of
equation~(\ref{fitCapel}). These are the uncertainties labeled II,
III, and~VI in Table~\ref{t:syserrors}. We assume that systematic
uncertainties can be added in quadrature. The final estimate of the
B-band zeropoint for local galaxies is then
\begin{equation}
\label{Zlocal}
  Z_{0.005} = 0.896 \pm 0.010 \> {\rm (random)}
                    \pm 0.049 \> {\rm (systematic)} ,
\end{equation}
where the value $\langle z \rangle = 0.005$ listed in the subscript is
the average redshift of the sample galaxies.


\newcommand{\figcapMLevol}{(a) $M/L$--$\sigma$ relation for 
the sample of local galaxies listed in Table~\ref{t:local} (same as in
the bottom right panel of Figure~\ref{f:local}). The solid line is the
best linear fit.  (b) The same quantities, but now for the sample of
intermediate redshift cluster galaxies modeled in Paper~I. The galaxies
are color-coded by the cluster in which they reside: CL 3C295 (blue),
CL 1601+4253 (red) or CL 0016+1609 (green). The cluster redshifts are
labeled on the plot. The dotted line is the best-fit for local
galaxies (as also shown in the left panel). The solid line has the
same slope, but was shifted to best fit the intermediate redshift
cluster galaxies. This shift quantifies the $M/L$ evolution with
redshift. The open symbol denotes the S0/Sb galaxy CL 3C295-568, which
was excluded from the analysis of $M/L$ evolution because it is not an
early-type galaxy (defined here as $T < 0$).\label{f:MLevol}}

\ifemulate
\begin{figure*}
\epsfxsize=0.85\hsize
\centerline{\epsfbox{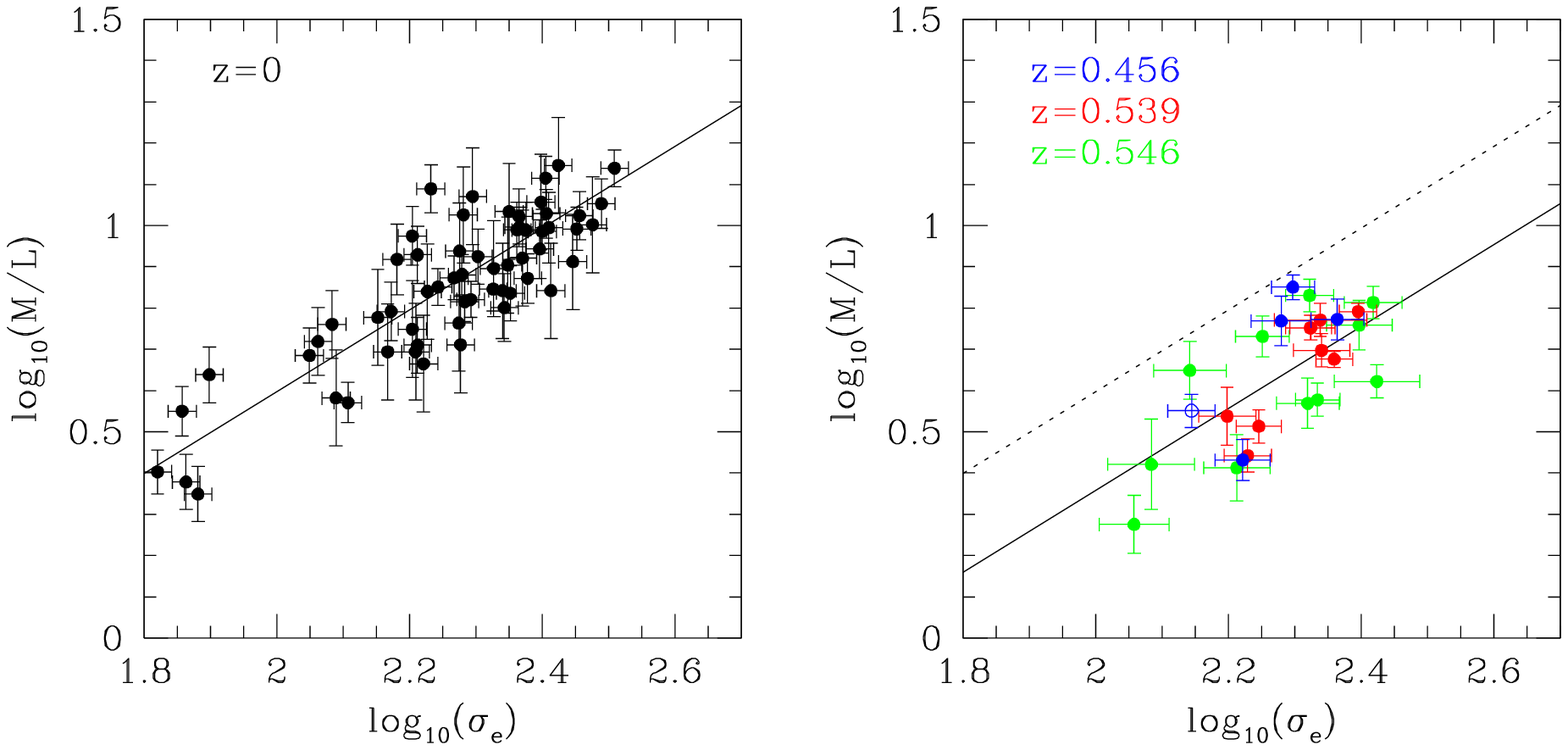}}
\figcaption{\figcapMLevol}
\end{figure*}
\fi


\section{Mass-to-Light Ratio Evolution}
\label{s:MLevolution}

\subsection{Intermediate-Redshift Cluster Galaxy Sample} 
\label{ss:sample}

The sample of Paper~I consists of 25 galaxies that reside in the
clusters CL 3C295, CL 0016+1609, and CL 1601+4253, at redshifts $z =
0.456$, $0.546$, and $0.539$, respectively (Dressler \& Gunn 1992;
Dressler \etal 1999). The clusters were selected based on their
visibility at the time of the Keck observations, and because they are
among the most S0 deficient clusters in the MORPHS sample (Dressler
\etal 1997). The latter criterion has little relevance for the results
discussed in the present paper, but was relevant for the discussion of
the rotation properties of the sample galaxies presented in
Paper~I. The MORPHS sample itself was not selected according to strict
criteria. The galaxy selection was largely constrained by the geometry
of the Keck/LRIS masks, and by the fact that sample galaxies should be
bright enough for spectroscopy. Priority was given to galaxies
classified from HST images as E or E/S0 by Smail \etal (1999). The
latest-type galaxy included in the sample was an S0/Sb galaxy. This
galaxy 3C295-568 was included for the specific purpose to see if
rotation could reliably measured (which is indeed the case, as
demonstrated in Paper~I).

The galaxy CL 3C295-2014 is the well-known AGN 3C295. There is the
possibility that this galaxy contains a central non-thermal point
source that could bias the analysis. However, the results for both the
light profile of this galaxy and its $M/L$ (see Paper~I and the
discussion in Section~\ref{ss:MLevol2I} below) do not provide any
evidence for deviations from the trends defined by the other galaxies
in the sample. We therefore retained CL 3C295-2014 in our sample and
did not treat it in any special way.

\subsection{Evolution of the $M/L$--$\sigma$ relation}
\label{ss:MLevol2I}

Figure~\ref{f:MLevol}b shows the $M/L$--$\sigma$ relation for our
sample of intermediate-redshift cluster galaxies, using the data from
Table~1 in Paper~I. We estimate the systematic uncertainties in
velocity dispersion estimates to be $\sim 5$\%, as in
Section~\ref{ss:dispersions} (i.e., $\Delta \log
\sigma_{\rm eff} = 0.021$). These uncertainties were added in
quadrature to the random uncertainties from Paper~I before plotting
and analysis. For comparison, the left panel of Figure~\ref{f:MLevol}a
shows the $M/L$--$\sigma$ relation for local galaxies, as in the
bottom right panel of Figure~\ref{f:local}.

The solid line in Figure~\ref{f:MLevol}b is the line with the fixed slope
$S = 0.992$ (as inferred from the local galaxy sample) that best fits
the intermediate redshift cluster galaxies. It has zeropoint
\begin{equation}
\label{Zdistant}
  Z_{0.528} = 0.657 \pm 0.022 \> {\rm (random)} 
                    \pm 0.049 \> {\rm (systematic)} .
\end{equation}
The value $\langle z \rangle = 0.528$ listed in the subscript is the
average redshift of the galaxies. The random uncertainty in the
zeropoint was determined as the ratio of the RMS residual with respect
to the fit and $\sqrt{N}$, where $N = 24$ is the number of
galaxies. This does not include the S0/Sb galaxy CL 3C295-568, which
was excluded from the analysis because it is not an early-type galaxy
(defined here as $T < 0$). By estimating the random zeropoint
uncertainty in this way we do not use the random uncertainties $\Delta
\log(M/L)$ in the individual $M/L$ measurements listed in
Paper~I. The average of these uncertainties is actually smaller than
the scatter around the best fit ($0.051$ vs.~$0.108$,
respectively). This may be due to the presence of additional random
uncertainties in addition to those propagated from the observed
kinematics, or it may be due to intrinsic scatter in the
$M/L$--$\sigma$ relation. The systematic uncertainty listed in
equation~(\ref{Zdistant}) is the quadrature sum of the relevant
uncertainties, namely I, III, IV, V and~VI in
Table~\ref{t:syserrors}. This assumes, in analogy with the local
sample, that the systematic uncertainty due to modeling limitations is
$\Delta_{\rm VI} = \pm 0.020$. This is reasonable, because the
two-integral modeling that we have used for the intermediate-redshift
cluster sample was very similar to that used by vdM91 and M98 for
local galaxies (and more generally, C06 found that two-integral models
do not yield strongly biased $M/L$ estimates as compared to more
sophisticated three-integral models).

The evolution of the $M/L$ between the two redshifts is obtained by
subtracting the zeropoints of the $M/L$--$\sigma$ relations given by
equations~(\ref{Zlocal}) and~(\ref{Zdistant}). This yields
\ifemulate
  \begin{eqnarray}
    \label{Zdiff}
    \delta && \log(M/L) \equiv Z_{\rm distant} - Z_{\rm local} \nonumber \\
           && = -0.239 \pm 0.024 \> {\rm (random)}
                       \pm 0.041 \> {\rm (systematic)} ,
  \end{eqnarray}
\else
  \begin{equation}
  \label{Zdiff}
    \delta \log(M/L) \equiv Z_{\rm distant} - Z_{\rm local}
                     = -0.239 \pm 0.024 \> {\rm (random)}
                            \pm 0.041 \> {\rm (systematic)} ,
  \end{equation}
\fi
where in this particular case the average redshifts for the distant
and local samples are $\langle z \rangle = 0.528$ and $0.005$,
respectively. The random uncertainty in $\delta \log(M/L)$ is simply
the quadrature sum of the random uncertainties in $Z_{0.528}$ and
$Z_{0.005}$. However, the systematic uncertainty is not the quadrature
sum of the systematic uncertainties in those quantities. That is
because source III of systematic uncertainty in
Table~\ref{t:syserrors} is common to both zeropoints, and therefore
drops out of their difference. So the systematic uncertainty listed in
equation~(\ref{Zdiff}) is the quadrature sum of the uncertainties I,
II, IV, V and~VI in Table~\ref{t:syserrors}. We include source VI only
once, because it was defined as a measure of the typical difference in
the results from different dynamical modeling approaches.

As discussed in Paper~I, the only parameter in our models that is not
generally constrained by the data is the inclination, or
alternatively, the intrinsic axial ratio $Q$. Equation~(\ref{Zdiff})
was derived from the results obtained in Paper~I for our ``standard
inclination'' models. These models use for each galaxy the most likely
inclination, given the observed projected axial ratio.  These models
have the correct average intrinsic axial ratio when averaged over a
large sample. For comparison we have also constructed two sets of
alternative models, namely models that are edge-on (yielding the
roundest possible intrinsic shape for each galaxy) and models that
have intrinsic axial $Q_{\rm min} = 0.4$ (which is approximately the
smallest intrinsic axial ratio found for early-type galaxies).  The
former yield an average $\delta \log(M/L)$ for the sample that is
lower by $-0.008$ than for the standard inclination models. The latter
yield a value that is higher by $0.053$. The assumptions that underly
these models are clearly unrealistic for the sample as a whole, and
either way do not change the result in equation~(\ref{Zdiff}) by much
more than the listed uncertainties. Nonetheless, these numbers give
some idea of the systematic uncertainty in the $M/L$ estimates {\it
for individual galaxies} due to the unknown inclinations. In other
words, variations in inclination between galaxies of the same
projected axial ratio do not affect the average correlation in
Figure~\ref{f:MLevol}b, but they do add to the scatter. However, the
induced scatter is too small to account for the observed scatter of
$0.108$ in $\log(M/L)$.

The change of $M/L$ with redshift is $\Delta \log(M/L) / \Delta z =
\delta \log(M/L) / (\langle z \rangle_{\rm distant} - 
\langle z \rangle_{\rm local})$. Equation~(\ref{Zdiff}) thus gives 
\ifemulate
  \begin{eqnarray}
  \label{Zevol}
    \Delta && \log(M/L) / \Delta z \nonumber \\
           && = -0.457 \pm 0.046 \> {\rm (random)}
                       \pm 0.078 \> {\rm (systematic)} .
  \end{eqnarray}
\else
  \begin{equation}
  \label{Zevol}
    \Delta \log(M/L) / \Delta z =
               -0.457 \pm 0.046 \> {\rm (random)}
                      \pm 0.078 \> {\rm (systematic)} .
  \end{equation}
\fi
So far we have treated all intermediate redshift cluster galaxies as a
single sample. It is of course also possible to calculate the
zeropoint evolution for each cluster individually. This yields $\delta
\log(M/L)$ (CL 3C295) $= -0.180 \pm 0.065$ (random),
$\Delta \log(M/L)$ (CL 1601+4253) $= -0.259 \pm 0.035$ (random), and
$\Delta \log(M/L)$ (CL 0016+1609) $= -0.241 \pm 0.024$ (random). These
values each have the same systematic uncertainty as listed for the
combined sample in equation~(\ref{Zdiff}).  The cluster {\rm CL 3C295}
has the smallest amount of evolution, as expected given its lower
redshift. However, the differences between the clusters are not really
significant given the random uncertainties. The clusters span a range
$\Delta z = \pm 0.045$ (see Section~\ref{ss:sample}), which implies an
expected variation $\Delta \log(M/L) = \pm 0.02$ on the basis of
equation~(\ref{Zevol}). This is smaller than the average random error
of $0.051$ in our $\log(M/L)$ measurements, and it is also smaller
than the scatter of $0.108$ around the linear fit in
Figure~\ref{f:MLevol}b. Correction for differential evolution between
galaxies in our sample at slightly different redshifts would therefore
not change any of the results. So it is justified to treat the
galaxies as a single sample at an average redshift $\langle z
\rangle = 0.528$. This approach has the advantage that it doesn't introduce
any prior knowledge about evolution into the analysis of the sample.


\newcommand{\tablecontentssysFPerrors}{
\tablecaption{Sources of systematic uncertainty that affect Fundamental Plane 
zeropoints\label{t:sysFPerrors}}
\tablehead{
\colhead{ID} & \colhead{$\Delta \zeta$} & 
\colhead{source of systematic uncertainty} & \colhead{Section} \\
}
\startdata
I   & $\pm 0.003$ & effect of uncertainties in $\Omega_{\rm m}$ and 
$\Omega_{\Lambda}$ at $z \approx 0.5$ & \ref{s:intro} \\
III & $\pm 0.037$ & accuracy of the Cepheid distance scale for local galaxies &
\ref{ss:dist} \\
IV  & $\pm 0.009$ & cosmic variance in $H_0$ on the scale $z \lta
0.1$ & \ref{ss:dist} \\
V   & $\pm 0.023$ & random uncertainty in $H_0$ due to finite sample sizes & 
\ref{ss:dist} \\
VII & $\pm 0.013$ & uncertainty in Hubble flow velocity of Coma & 
\ref{ss:MLevolFP} \\
VIII& $\pm 0.020$ & difference in FP zeropoint determinations from 
different authors, at fixed distance & \ref{ss:MLevolFP} \\
IX  & $\pm 0.018$ & cosmic variance in $H_0$ on the scale $z \lta
0.025$ & \ref{ss:MLevolFP} \\
\enddata}

\newcommand{\tablecommsysFPerrors}{Column~(1) lists the roman 
numeral with which a particular systematic uncertainty is referred to
in the text. Column~(2) lists the size of the systematic uncertainty
in the Fundamental Plane zeropoint $\zeta$. The corresponding
uncertainty in mass-to-light ratio is $\Delta
\log(M/L) = \Delta \zeta / 0.83$. Column~(3) lists the source of the 
systematic uncertainty. Column~(4) lists the number of the section in which
the uncertainty is discussed.}

\ifemulate
\begin{deluxetable*}{lclc}
\tabletypesize{\footnotesize}
\tablecontentssysFPerrors
\tablecomments{\tablecommsysFPerrors}
\end{deluxetable*}
\fi


\subsection{Comparison to Fundamental Plane evolution}
\label{ss:MLevolFP}

In vDvdM06 we studied the evolution of the FP of the sample
clusters. The analysis used the relation
\begin{equation}
\label{FPrel}
  \log r_{\rm eff} = {\rm FP} - \zeta ,  
\end{equation}
where 
\begin{equation}
\label{FPdef}
  {\rm FP} \equiv 1.20 \log \sigma_{\rm ap} - 0.83 \log I_{\rm eff} .
\end{equation}
Here $\sigma_{\rm ap}$ is the dispersion in an aperture radius that
spans $1.7''$ at the distance of the Coma cluster (this follows
Jorgensen \etal 1995b) and $I_{\rm eff}$ is the average rest-frame
$B$-band surface brightness inside $r_{\rm eff}$. From the shift in
the zeropoint $\zeta$ of the relation with respect to the Coma cluster
one infers an $M/L$ evolution of
\ifemulate
  \begin{eqnarray}
  \label{FPdiff}
    \delta && \log(M/L) \equiv (\zeta_{\rm distant} - \zeta_{\rm coma}) 
                               / 0.83 \nonumber \\
           && = -0.268 \pm 0.025 \> {\rm (random)}
                       \pm 0.036 \> {\rm (systematic)} .
  \end{eqnarray}
\else
  \begin{equation}
  \label{FPdiff}
    \delta \log(M/L) \equiv (\zeta_{\rm distant} - \zeta_{\rm coma}) / 0.83 
                     = -0.268 \pm 0.025 \> {\rm (random)}
                              \pm 0.036 \> {\rm (systematic)} .
  \end{equation}
\fi
This result uses only the same subset of $N=17$ galaxies that were
included in the FP analysis in vDvdM06 (we discuss this selection
further in Section~\ref{s:slopeevol} below). It treats all the
intermediate-redshift cluster galaxies as a single sample, for
consistency with Section~\ref{ss:MLevol2I}. The average redshift of
this sample is $\langle z \rangle = 0.531$. The random uncertainty in
$\delta \log(M/L)$ is $1/0.83$ times the quadrature sum of the random
uncertainties in the FP zeropoints for Coma ($0.011$) and the
intermediate-redshift cluster galaxy sample ($0.023$). For Coma we
used the data for galaxies that have $B$-band photometry listed in
Jorgensen \etal (1995a) and velocity dispersions listed in Jorgensen
\etal (1995b). For each sample the random uncertainty was estimated as
before as the ratio of the RMS residual with respect to the fit and
$\sqrt{N}$.

The systematic uncertainty in equation~(\ref{FPdiff}) is due to a
combination of the systematic uncertainties that affect $\zeta_{\rm
coma}$ and $\zeta_{\rm distant}$. We summarize these uncertainties in
Table~\ref{t:sysFPerrors}. Sources that were already encountered
previously are listed with the same roman numeral as in
Table~\ref{t:syserrors}. Consider first the Coma FP zeropoint
$\zeta_{\rm coma}$. Distance uncertainties affect $\log r_{\rm eff}$,
and therefore the zeropoint of the FP relation. Distances in turn are
estimated as the ratio of the Hubble flow velocity $v_{\rm flow}$ and
the Hubble constant $H_0$. Uncertainties in both of these introduce FP
zeropoint uncertainties. We have used in our analysis one of the most
recent flow velocity estimates for Coma, namely $v_{\rm flow} = 7376
\pm 223 \kms$ from the SMAC survey (Smith, Lucey \& Hudson 2006). This is 
consistent with several earlier results. For example, F89 obtained
$v_{\rm flow} = 7461 \pm 273 \kms$ and Colless \etal (2001) obtained
$v_{\rm flow} = 7238 \pm 302 \kms$. The uncertainties in all these
results are dominated by the modeling uncertainty in the peculiar
velocity $v_{\rm pec} \equiv v_{\rm obs} - v_{\rm flow}$ of the Coma
cluster, where $v_{\rm obs}$ is the observed systematic velocity. The
uncertainty in $v_{\rm flow}$ introduces an uncertainty of $\Delta
\zeta_{\rm VII} = \pm 0.013$ in $\zeta$. The determination of the
Hubble flow velocity also has a small dependence on the cosmological
parameters $\Omega_{\rm m}$, $\Omega_{\Lambda}$. However, at the
distance of Coma this dependence is small enough that the resulting
uncertainties can be neglected.  Uncertainties in $H_0$ include the
uncertainties III and V in Table~\ref{t:syserrors}, which introduce FP
zeropoint uncertainties $\Delta \zeta_{\rm III} = \pm 0.037$ and
$\zeta_{\rm III} = \pm 0.023$. The last systematic uncertainty that
affects $\zeta_{\rm coma}$ stems from the fact that different authors
get slightly different zeropoints, even if they assume exactly the
same distance. Based on our own analysis of various literature
sources, as well as the detailed analysis of Hudson \etal (2001), we
estimate this systematic zeropoint uncertainty to be $ \Delta
\zeta_{\rm VIII} = \pm 0.020$. Addition in quadrature of the 
uncertainties III, V, VII, and~VIII yields a final systematic
uncertainty in $\zeta_{\rm coma}$ of $\pm 0.050$. The FP zeropoint
$\zeta_{\rm distant}$ of the intermediate-redshift cluster galaxy
sample shares the systematic uncertainties III, V, and~VIII with
Coma. It also is subject to uncertainty I in Table~\ref{t:syserrors}
that results from uncertainties in $\Omega_{\rm m}$ and
$\Omega_{\Lambda}$, and uncertainty IV that results from cosmic
variance in $H_0$. Addition in quadrature yields a final systematic
uncertainty in $\zeta_{\rm distant}$ of $\pm 0.049$. To study $M/L$
evolution we are now interested in the difference $\zeta_{\rm coma} -
\zeta_{\rm distant}$. The uncertainties III, IV, and~V in
Table~\ref{t:sysFPerrors} drop out of this difference. However, the
difference is subject to uncertainties I, VII and~VIII. We include
source VIII only once, because it was defined as a measure of the
typical difference in zeropoint between different authors. There is
also a new uncertainty due to cosmic variance in $H_0$ that is similar
to source IV. However, the relevant scale is now the distance of the
Coma cluster ($z = 0.025$), and not the scales $z
\lta 0.1$ over which Type Ia supernovae have been used to calibrate
$H_0$. Based on the discussion in Section~8.6 of Freedman
\etal (2001) we estimate this systematic component to be less than
4\%. This corresponds to a FP zeropoint difference $\Delta
\zeta_{\rm IX} = \pm 0.018$. Addition in quadrature yields a
final systematic uncertainty in $\zeta_{\rm coma} - \zeta_{\rm
distant}$ of $\pm 0.030$. Division by $0.83$ yields the systematic
uncertainty in $\delta \log(M/L)$ listed in equation~(\ref{FPdiff}).

The change with redshift implied by the FP zeropoint evolution listed
in equation~(\ref{FPdiff}) is $\Delta \log(M/L) / \Delta z = \delta
\log(M/L) / (\langle z \rangle_{\rm distant} - \langle z 
\rangle_{\rm local})$, which gives
\ifemulate
  \begin{eqnarray}
  \label{ZevolFP}
    \Delta && \log(M/L) / \Delta z = \nonumber \\
           && -0.529 \pm 0.049 \> {\rm (random)}
                     \pm 0.071 \> {\rm (systematic)} .
  \end{eqnarray}
\else
  \begin{equation}
  \label{ZevolFP}
    \Delta \log(M/L) / \Delta z =
               -0.529 \pm 0.049 \> {\rm (random)}
                      \pm 0.071 \> {\rm (systematic)} .
  \end{equation}
\fi
This FP result agrees with that of vDvdM06, which found $\Delta
\log(M/L) / \Delta z = -0.555 \pm 0.042 \> {\rm (random)}$ from a FP study of
a larger sample of clusters that included the three clusters studied
here.

The difference between the $M/L$ evolution derived from the
$M/L$--$\sigma$ relation and dynamical modeling of internal
kinematics, as reported in equation~(\ref{Zevol}), and that derived
from FP evolution, as reported in equation~(\ref{ZevolFP}), is
\ifemulate
  \begin{eqnarray}
  \label{DynFPcomp}
    [ \Delta && \log(M/L)_{\rm dyn} - \Delta \log(M/L)_{\rm FP} ] / \Delta z =
                \nonumber \\
             && 0.072 \pm 0.067 \> {\rm (random)}
                      \pm 0.097 \> {\rm (systematic)} .
  \end{eqnarray}
\else
  \begin{equation}
  \label{DynFPcomp}
    [ \Delta \log(M/L)_{\rm dyn} - \Delta \log(M/L)_{\rm FP} ] / \Delta z =
                0.072 \pm 0.067 \> {\rm (random)}
                      \pm 0.097 \> {\rm (systematic)} .
  \end{equation}
\fi
The listed random uncertainty is the quadrature sum of the random
uncertainties in equations~(\ref{Zevol}) and~(\ref{ZevolFP}). This
might be a slight overestimate, because it ignores possible
correlations between the residuals from both methods. The listed
systematic uncertainty is the quadrature sum of all those
uncertainties that do not drop out of the difference under
consideration. This includes the systematic uncertainties~II, V,
and~VI in Table~\ref{t:syserrors} and VII and VIII in
Table~\ref{t:sysFPerrors}. There also is a systematic uncertainty due
to the cosmic variance in $H_0$ when measured on scales of $z \lta
0.025$ and $z \lta 0.1$, respectively. Based on the discussion in
Section~8.6 of Freedman \etal (2001) we estimate this uncertainty to
be $\Delta \log(M/L) = \pm 0.009$.

The upshot of equation~(\ref{DynFPcomp}) is that the $M/L$ evolution
derived here from detailed dynamical models is consistent with that
derived from FP analysis of global parameters. The present paper
therefore supports the conclusions drawn in vDvdM06 (and summarized in
Section~\ref{s:intro} of the present paper) about the star formation
epoch of early type galaxies. 


\newcommand{\figcapMLres}{$\log (M/L)$ residuals with respect to
the solid line in Figure~\ref{f:MLevol}b. There is no significant
correlation with $\log \sigma_{\rm eff}$. So there is no evidence for
evolution of the slope of the $M/L$--$\sigma$ relation out to $z
\approx 0.5$. Each datapoint is an individual galaxy in the sample 
of intermediate redshift cluster galaxies, with point types the same
as in Figure~\ref{f:MLevol}.\label{f:MLres}}

\ifemulate
\begin{figure}
\epsfxsize=0.93\hsize
\centerline{\epsfbox{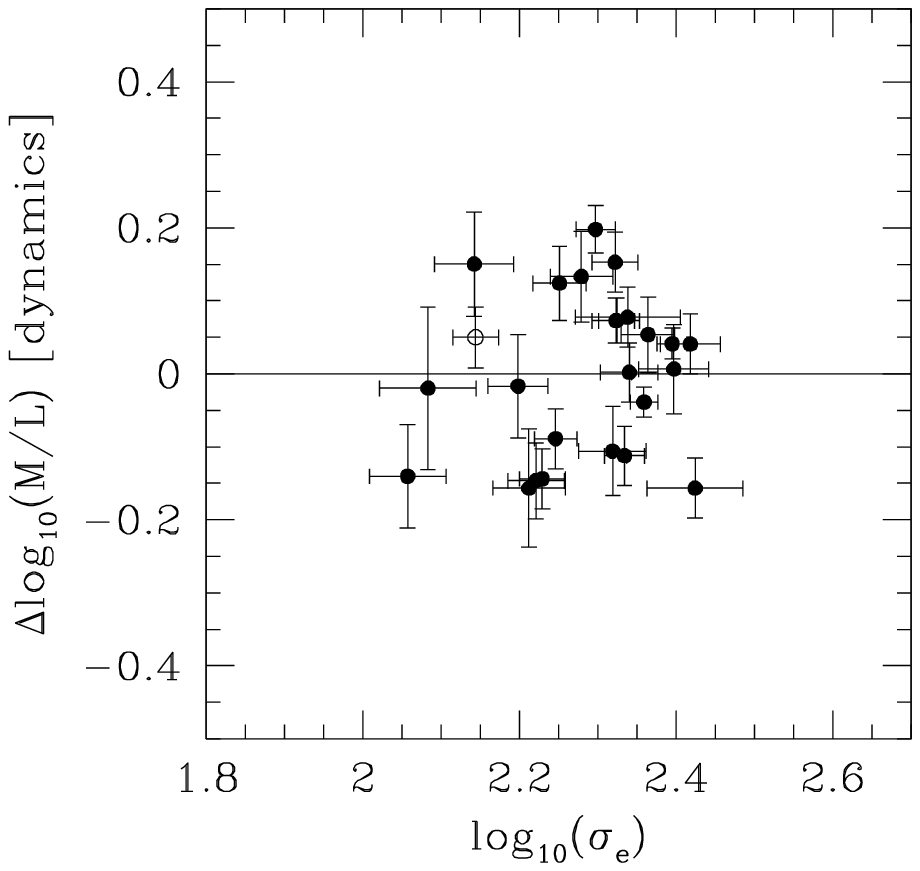}}
\figcaption{\figcapMLres}
\end{figure}
\fi


\section{Dependence of $M/L$ evolution on $\sigma$}
\label{s:slopeevol}

Figure~\ref{f:MLres} shows the residuals with respect to the
$M/L$--$\sigma$ relation (i.e., with respect to the solid line in
Figure~\ref{f:MLevol}b) for the sample of intermediate redshift
cluster galaxies. The $\log (M/L)$ residuals do not show a significant
correlation with $\log \sigma_{\rm eff}$. For example, the average
residual for galaxies with $\sigma_{\rm eff} < 200\kms$ is $-0.010 \pm
0.039$, while the average residual for galaxies with $\sigma_{\rm eff}
> 200\kms$ is $0.008 \pm 0.024$. These values are consistent at the
1-$\sigma$ level. This implies that the analysis yields the same
evolution $\delta \log(M/L)$ (eq.~[\ref{Zdiff}]) independent of
whether or not galaxies with low $\sigma$ are excluded from the
statistics. Phrased differently, there is no evidence for evolution of
the slope of the $M/L$--$\sigma$ relation out to $z \approx 0.5$. This
can also be shown explicitly. A straight line fit to the data in
Figure~\ref{f:MLevol}b yields slope $S = 1.117 \pm 0.113$, as compared
to $S = 0.992 \pm 0.054$ for the sample of local galaxies in
Figure~\ref{f:MLevol}a.  Again, these values are consistent at the
1-$\sigma$ level.

The relative constancy of the slope of the $M/L$--$\sigma$ relation is
consistent with some studies of FP evolution. For example, Kelson
\etal (2000) find no statistically significant change in the tilt of the
FP between the local Universe and a cluster at $z=0.33$. However, many
other FP studies, mostly towards higher redshifts, have recently found
that the FP tilt does evolve (e.g., Wuyts \etal 2004; Treu \etal
2005a,b; van der Wel \etal 2004, 2005; di Serego Alighieri \etal 2005; and
Jorgensen \etal 2006). In fact, we showed in vDvdM06 that the FP
residuals for the same sample analyzed here show more evolution for
galaxies with $\sigma \lta 200 \kms$ (or similarly, $M\lta 10^{11}
\Msun$) than for galaxies with $\sigma \gta 200 \kms$ (or similarly,
$M\gta 10^{11} \Msun$). Based on these results, the FP analysis in
vDvdM06 was restricted to galaxies with $M > 10^{11}
\Msun$. The same selection was applied in the derivation of
equation~(\ref{ZevolFP}). With this FP selection criterion, the
inferred $M/L$ evolution agrees with that derived from the dynamical
models presented here (as quantified by
eq.~[\ref{DynFPcomp}]). However, without this FP selection criterion
the agreement is worse.

This issue is further illustrated in Figure~\ref{f:DynFP}. It shows
for each galaxy the difference between the evolution in $\log(M/L)$
inferred either from the dynamically inferred $M/L$--$\sigma$ relation
or from the FP. The differences are plotted as a function of $\log
\sigma_{\rm eff}$. There is a clear trend with $\sigma_{\rm eff}$. The
results from the two methods agree only when the comparison is
restricted to galaxies with $\sigma \gta 200 \kms$ (or similarly,
$M\gta 10^{11} \Msun$). It is shown by equation~(\ref{DynFPcomp}) that
the difference between the two methods might be affected by various
systematic uncertainties. However, these uncertainties are mostly
related to uncertainties in distance scales. They cannot introduce a
dependence on $\log \sigma_{\rm eff}$. Therefore, the trend in
Figure~\ref{f:DynFP} must have a different origin. We continue in the
following subsections by exploring various possible explanations.
This is an important issue, because differences in $M/L$ evolution
between low-mass and high-mass galaxies are generally interpreted as
due to differences in stellar population age. Such age differences
have a direct bearing on our understanding of galaxy formation.

\subsection{Structure Evolution}
\label{ss:strucevol}

If galaxies do not change their structure over time then any evolution
in either the FP or the $M/L$--$\sigma$ relation must be due to $M/L$
evolution. In this case, a study of the evolution of these two
relations always implies the same (correct) $M/L$ evolution. However,
this equivalence ceases to exist when the structure of galaxies
evolves with time.

As a simple illustration of the possible impact of structure
evolution, consider the example in which some process changes the
$r_{\rm eff}$ of a galaxy while $M/L$ and $\sigma_{\rm eff}$ remain
constant. This will obviously not move the galaxy with respect to the
$M/L$--$\sigma$ relation. However, the galaxy will move with respect
to the FP. The virial theorem dictates that
\begin{equation}
\label{virial}
   (M/L)_{\rm vir} = K \sigma_{\rm eff}^2 / (2 \pi G r_{\rm eff}
                     I_{\rm eff}) ,
\end{equation}
where $G$ is the gravitational constant and $K$ the structure
constant, a scalar that depends on the structure of the galaxy (for
example, a spherical model with a de Vaucouleurs $R^{1/4}$ surface
brightness profile has $K=5.95$). If we assume in this example that
the galaxy changes its structure homologously, then $K$ will remain
constant. Equation~(\ref{virial}) then implies that $\Delta \log
I_{\rm eff} = - \Delta \log r_{\rm eff}$. Therefore, at its new
$r_{\rm eff}$ the galaxy will be offset from the FP by an amount
$\Delta \zeta = -0.17 \Delta \log r_{\rm eff}$. The customary
assumptions for interpreting FP evolution (see e.g.,
eq.~[\ref{FPdiff}]) would then incorrectly suggest a mass-to-light
ratio evolution of $\Delta \log(M/L) = -0.20 \Delta \log r_{\rm eff}$.


\newcommand{\figcapDynFP}{Difference between 
the evolution in $\log (M/L)$ as calculated either from dynamical
models and the $M/L$--$\sigma$ relation, or from the FP. The inferred
evolution agrees for dispersions $\sigma_{\rm eff} \gta 200 \kms$
(i.e., $\log \sigma_{\rm eff} \gta 2.30$; or similarly, $M\gta 10^{11}
\Msun$). However, at lower dispersions the FP suggests more evolution
than do the dynamical $M/L$ measurements. Each datapoint is an
individual galaxy in the sample of intermediate redshift cluster
galaxies, with point types the same as in
Figure~\ref{f:MLevol}.\label{f:DynFP}}

\ifemulate
\begin{figure}
\epsfxsize=0.8\hsize
\centerline{\epsfbox{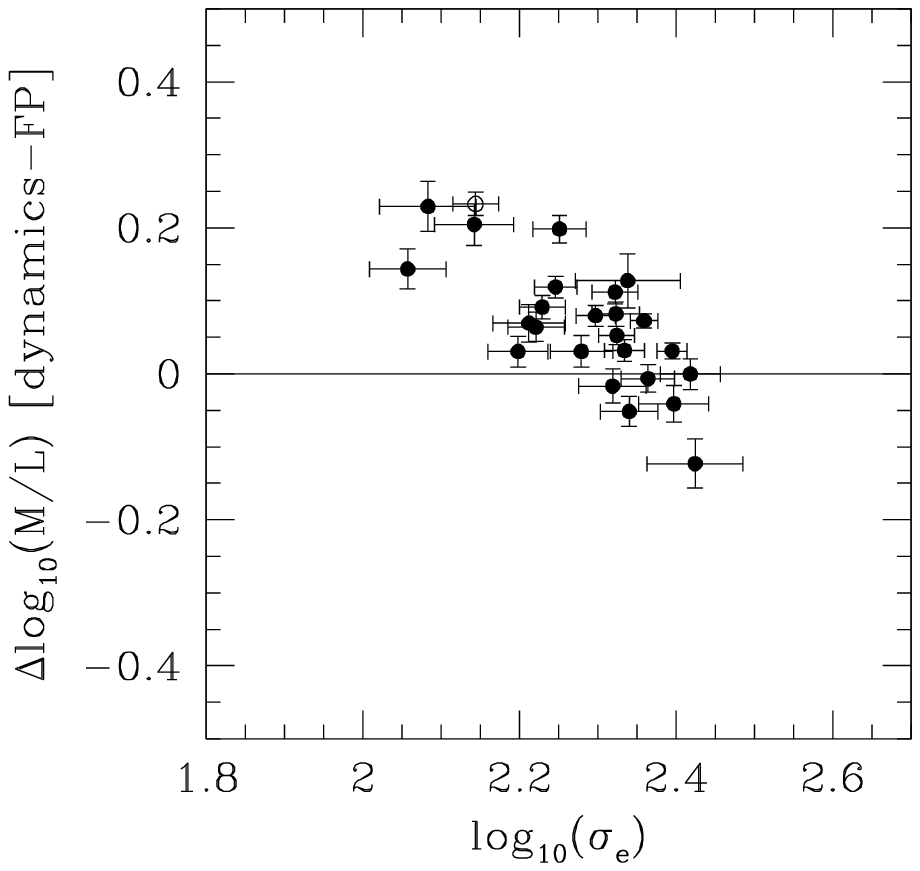}}
\figcaption{\figcapDynFP}
\end{figure}
\fi


It is equally possible to construct examples in which FP evolution
properly measures $M/L$ evolution, whereas $M/L$--$\sigma$ evolution
does not. To illustrate this, consider the situation in which due to
some process a galaxy homologously changes its $r_{\rm eff}$,
$\sigma_{\rm eff}$ and $I_{\rm eff}$ while $M/L$ remains constant. If
$\Delta \log r_{\rm eff} = -2.71 \Delta \log
\sigma_{\rm eff}$ and $\Delta \log I_{\rm eff} = 4.71 \Delta \log
\sigma_{\rm eff}$, then the galaxy will continue to fall on the
edge-on projection of the FP. The customary assumptions for
interpreting FP evolution would then correctly suggest that there was
no change in $M/L$. However, the galaxy will now be offset from the
$M/L$--$\sigma$ relation by an amount $\Delta \log(M/L) = 0.37 \Delta
\log r_{\rm eff}$.

These examples show that whenever there is evolution in any of the
structural quantities $\sigma_{\rm eff}$, $r_{\rm eff}$, $\Sigma_{\rm
eff} \equiv I_{\rm eff} (M/L)$, or $K$, then the $M/L$ evolution
inferred from the FP and from the $M/L$--$\sigma$ relation will not
generally agree with each other. One plausible explanation for the
trend in Figure~\ref{f:DynFP} is therefore that galaxies undergo
structural evolution with time, and that this evolution is different
for galaxies of low and high $\sigma_{\rm eff}$.

The assessment of the importance of this effect is complicated by the
fact that our analysis of the evolution of the $M/L$--$\sigma$
evolution has been based on rather different data than our analysis of
the FP evolution. The former used spatially resolved brightness
profiles and kinematics, whereas the latter used only the
characteristic quantities $r_{\rm eff}$, $I_{\rm eff}$, and
$\sigma_{\rm eff}$. A somewhat cleaner comparison can therefore be
made through an analysis of $M/L$--$\sigma$ evolution based on virial
estimates from equation~(\ref{virial}), which uses the same three
characteristic quantities as FP studies. To this end, we first
calculated the values of $(M/L)_{\rm vir}$ for the sample of Coma
galaxies studied by Jorgensen \etal (1995a,b). The values thus
obtained were fit by a straight line of the form given by
equation~(\ref{fitCapel}) which yields parameters $Z_{\rm coma} = \log
K + 0.163 \pm 0.015$ and $S_{\rm coma} = 0.887 \pm 0.095$. We then
calculated $(M/L)_{\rm vir}$ for all of the intermediate-redshift
cluster galaxies. The evolution for each galaxy was calculated by
comparison to the $(M/L)_{\rm vir}$--$\sigma$ relation for Coma.


\newcommand{\figcapvirial}{Difference between 
the evolution in $\log (M/L)$ as calculated either from the virial
theorem and the $M/L$--$\sigma$ relation, or from the FP. The two
methods do not yield the same results, despite the fact that both
analyses are based on exactly the same global properties ($\sigma_{\rm
eff}$, $r_{\rm eff}$, and $I_{\rm eff}$) and use exactly the same
local comparison sample (Coma). There is a trend with $\sigma_{\rm
eff}$ in the same direction as in Figure~\ref{f:DynFP}. Each
datapoint is an individual galaxy in the sample of intermediate
redshift cluster galaxies, with point types the same as in
Figure~\ref{f:MLevol}.\label{f:virial}}

\ifemulate
\begin{figure}
\epsfxsize=0.8\hsize
\centerline{\epsfbox{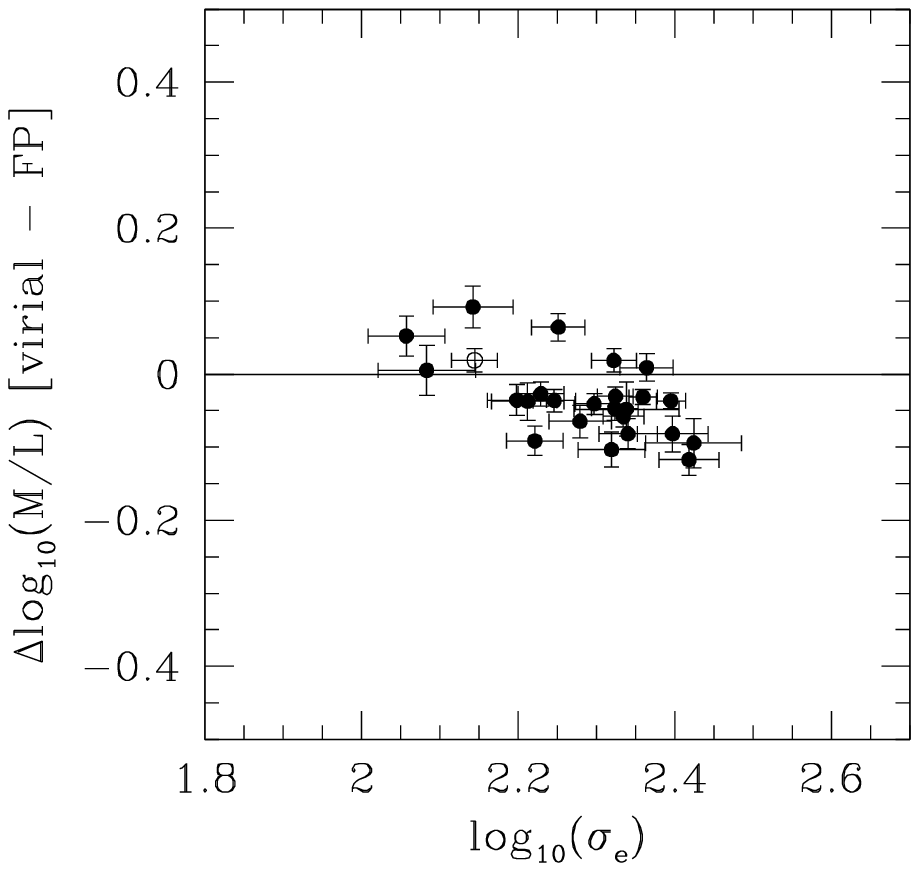}}
\figcaption{\figcapvirial}
\end{figure}
\fi


Figure~\ref{f:virial} shows for each galaxy the difference between the
evolution thus determined and the evolution determined from the FP
evolution. By contrast to Figure~\ref{f:DynFP}, both measures of
evolution are now inferred entirely from the same global properties
($\sigma_{\rm eff}$, $r_{\rm eff}$, and $I_{\rm eff}$), and use
exactly the same local comparison sample. However, there is a still a
trend in the inferred $M/L$ evolution with $\sigma_{\rm eff}$, in the
same sense as in Figure~\ref{f:DynFP}. This supports the view that
structural quantities in addition to $M/L$ are probably evolving with
time. Note that it is not possible to identify whether a more accurate
estimate of $M/L$ evolution is obtained by studying FP evolution or
$M/L$--$\sigma$ evolution. For this one would need to know exactly how
the structural properties of individual elliptical galaxies change with
redshift, which is not well constrained observationally.

\subsection{Rotation}
\label{ss:rotation}

The trend in Figure~\ref{f:virial} is not as steep as that in
Figure~\ref{f:DynFP}. This means that part of the trend in
Figure~\ref{f:DynFP} must be due to differences in evolution between
our $M/L$ values from dynamical modeling and the $(M/L)_{\rm vir}$
values determined from the virial theorem. Such differences are not
necessarily unexpected, given that our models ought to be more
accurate. They account for rotation and homological differences,
between galaxies and as a function of redshift, whereas the virial
analysis does not.


\newcommand{\figcaprot}{Difference between
the evolution in $\log (M/L)$ as calculated from the $M/L$--$\sigma$
relation using either dynamical models or the virial theorem. The
difference is shown as a function of the rotation rate $k$ determined
in Paper~I. For rapidly rotating galaxies the $M/L$ evolution inferred
from the dynamical models is less than that implied by the virial
estimates. Each datapoint is an individual galaxy in the sample of
intermediate redshift cluster galaxies, with point types the same as
in Figure~\ref{f:MLevol}. Only those 15 galaxies are shown for which
the rotation rate $k$ could be reliably determined.\label{f:rot}}

\ifemulate
\begin{figure}
\epsfxsize=0.8\hsize
\centerline{\epsfbox{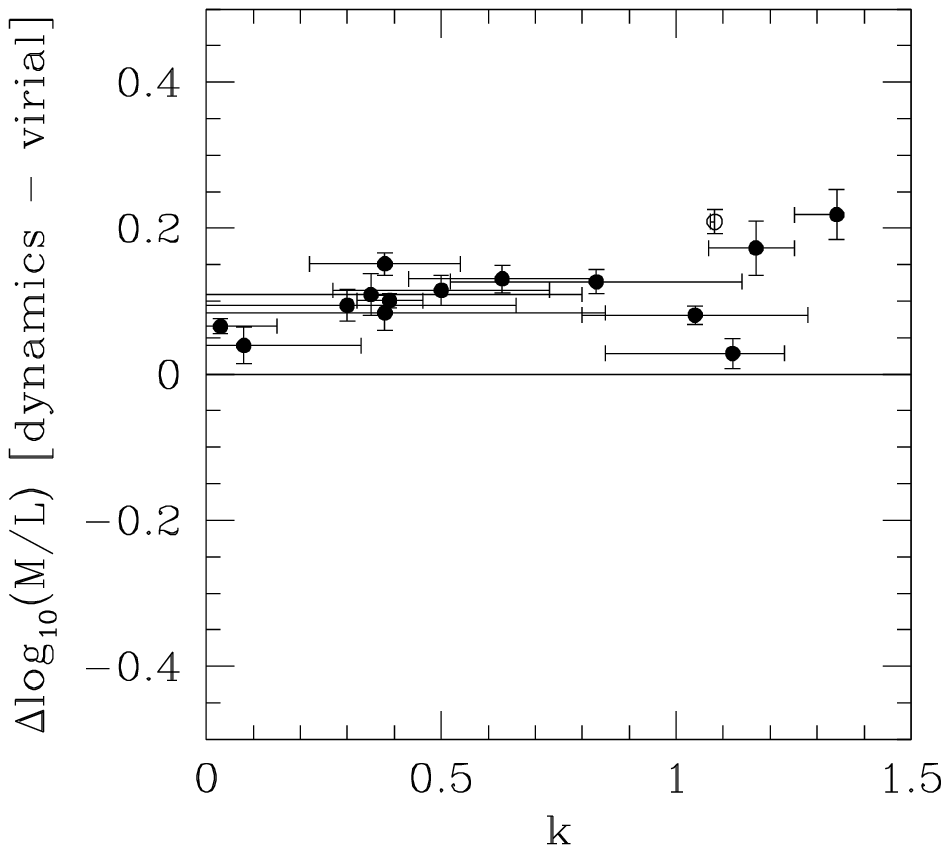}}
\figcaption{\figcaprot}
\end{figure}
\fi


In Paper~I we determined for each galaxy the normalized rotation
measure $k$, which is similar to the quantity $(v/\sigma)^*$ that is
often used for local galaxies. For $k=0$ the galaxy is non-rotating,
whereas for $|k|=1$ the velocity dispersion tensor is isotropic and
the galaxy is a so-called ``oblate isotropic rotator''. The quantity
$k$ is reasonably well determined for the 15 galaxies in the sample
for which a slit was placed within $45^{\circ}$ of the major axis, and
which have projected ellipticity $\epsilon > 0.10$ (these galaxies are
marked with an asterisk in Table~1 of Paper~I). Figure~\ref{f:rot}
shows for these 15 galaxies the differences between the following two
quantities: (a) the evolution inferred from our dynamical models and
the local comparison sample described in Section~\ref{s:local}; and
(b) the evolution inferred from the $(M/L)_{\rm vir}$ estimates using
Coma as comparison sample. Since both measures of evolution are based
on an underlying relation between $M/L$ and $\sigma$, any effects
introduced by structure evolution, such as those described in
Section~\ref{ss:strucevol}, are now removed from the comparison. The
differences thus calculated (which are equal to the the differences
between the residuals shown in Figures~\ref{f:DynFP}
and~\ref{f:virial}) are shown as a function of the rotation parameter
$k$. There is a significant trend for the residuals to be larger for
galaxies that have more rotational support. For example, the quantity
$\Delta \log(M/L)_{\rm dyn} - \Delta \log(M/L)_{\rm vir}$ has an
average value of $0.095 \pm 0.011$ for galaxies with $k < 0.6$ and
$0.138 \pm 0.026$ for galaxies with $k > 0.6$.  By contrast, we have
found no correlation with the observed axial ratio of each galaxy.

The dynamical models that we have used here explicitly account for the
observed rotation of each galaxy, whereas studies based entirely on
$\sigma_{\rm eff}$ do not. This causes $(M/L)_{\rm vir}$ or the FP to
systematically underestimate the $M/L$ of rapidly rotating
galaxies. It is not immediately obvious though that the inferred $M/L$
{\it evolution} would be impacted by this. Studies of FP evolution or
$(M/L)_{\rm vir}$--$\sigma$ evolution ignore rotation both in the
local Universe and at intermediate redshift, so any bias might cancel
out when different redshifts are compared. On the other hand,
observations in the local universe generally use small apertures that
cover only the central part of the galaxy, whereas observations of
distant galaxies generally use apertures that cover most of the
galaxy. To enable meaningful comparison, observations at different
redshifts are generally transformed to a common aperture size using
transformations that themselves do not explicitly account for rotation
(e.g., Jorgensen \etal 1995b). Therefore, observations at different
redshifts may be impacted by rotation in different manner. So any bias
introduced through the neglect of rotation in FP or $(M/L)_{\rm vir}$
calculations may not cancel out when samples at different redshifts
are compared. This may bias the inferred evolution and could therefore
affect the trend in Figure~\ref{f:rot}. The fact that the trend in
Figure~\ref{f:DynFP} is steeper than that in Figure~\ref{f:virial} is
related to the fact that galaxies of low $\sigma_{\rm eff}$ tend to
have more rotational support than those of high $\sigma_{\rm eff}$
(both in the local universe and in the sample of intermediate-redshift
cluster galaxies, see Paper~I).

Figure~\ref{f:rot} shows that even at $k \approx 0$ there is a
difference of $\sim 0.05$ dex between the evolution inferred from the
dynamical and virial $M/L$ estimates. The significance of this offset
is low, given the systematic error quoted in
equation~(\ref{DynFPcomp}). The offset could be due to (a combination
of) several different effects. For example: (a) systematic
uncertainties in the relative distance scale between our local
comparison sample of Section~\ref{s:local} and Coma; (b) systematic
errors in the conversions of observed velocity dispersions in some
fixed aperture to $\sigma_{\rm eff}$; or (c) redshift evolution of the
structure constant $K$.


\newcommand{\figcapMLrotresids}{$\log (M/L)$ residuals for the 
intermediate redshift cluster galaxies with respect to the solid line
in Figure~\ref{f:MLevol}b. (a) Residuals versus rotation rate $k$ for
the 15 galaxies with reliable $k$ determinations. (b) Residuals versus
Hubble $T$ type from Smail \etal (1997) for the full sample; $-5$,
$-4$, $-3$, $-2$, and $0$ correspond to E, E/S0, S0/E, S0, and S0/Sa,
respectively. There is no obvious trend in either panel that might
have supported the view that more rapidly rotating or later-type
galaxies formed the bulk of their stars more
recently.\label{f:MLrotresids}}

\ifemulate
\begin{figure*}
\epsfxsize=0.8\hsize
\centerline{\epsfbox{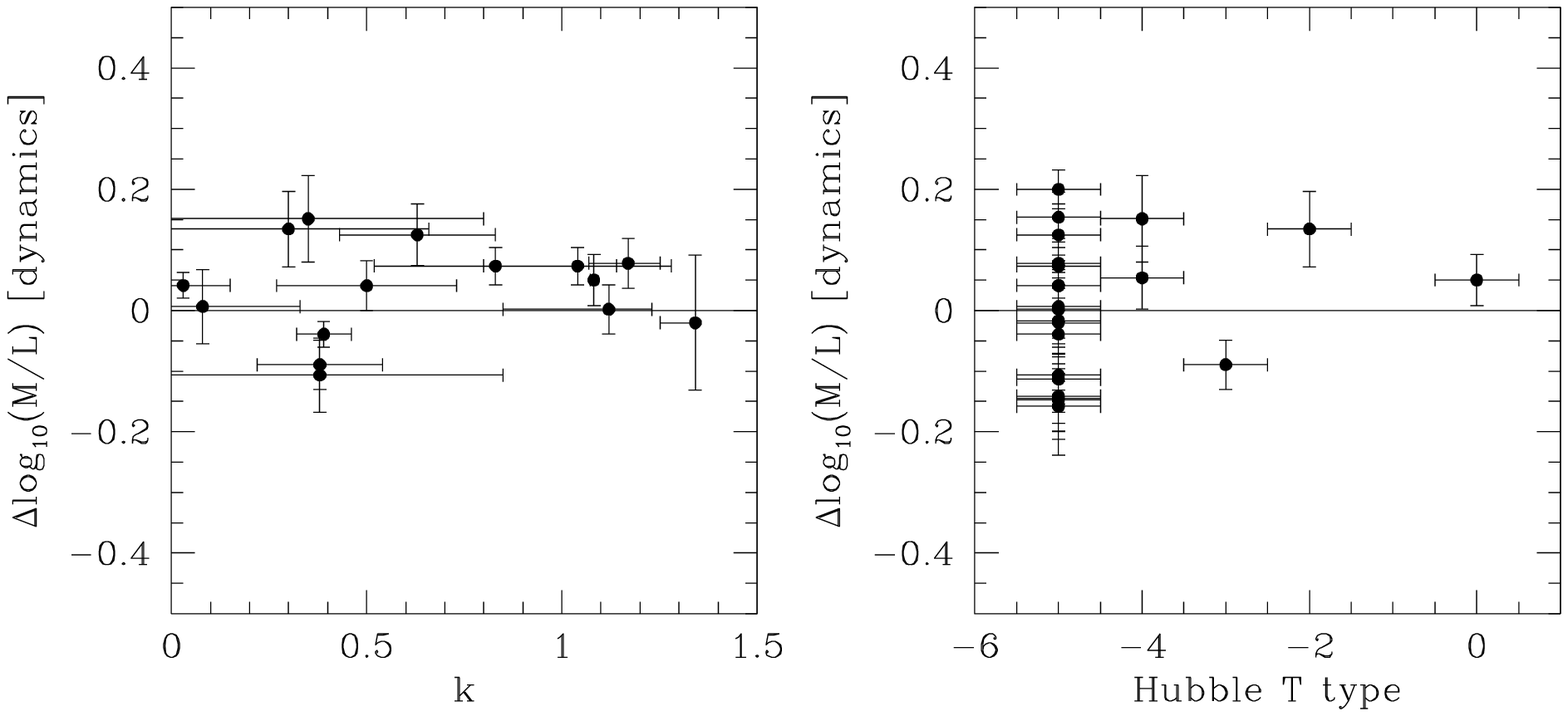}}
\figcaption{\figcapMLrotresids}
\end{figure*}
\fi


On a separate note, we do not find a correlation between the residuals
from the $M/L$--$\sigma$ relation inferred from our dynamical models
(i.e., the offset of the points in Figure~\ref{f:MLevol}b from the
solid straight line in that figure) and either the rotation parameter
$k$ or the morphological types from Smail \etal (1997). This is shown
in Figure~\ref{f:MLrotresids}. We discussed in Paper~I that some of
the most rapidly rotating galaxies in the sample could be
misclassified S0 galaxies. Such galaxies, or other galaxies of
intermediate and late Hubble types, might have recently transformed
from star-forming field galaxies. If so, one would expect their $M/L$
ratios to be smaller than for the true elliptical galaxies, which may
have formed the bulk of their stars longer
ago. Figure~\ref{f:MLrotresids} shows no obvious trends that would
support this view. However, a sample with a larger number of S0 and
later-type galaxies would obviously be more suited to address this
issue.

\subsection{Other Systematic Effects}
\label{ss:other}

The slope of the $M/L$--$\sigma$ relation does not change with
redshift for our sample, despite the fact that the FP tilt does evolve
with redshift. We have shown that this can be plausibly attributed to
a combination of two effects: (a) evolution in structural properties;
and (b) the neglect of rotational support in studies of FP evolution.
Nonetheless, there are other systematic effects that may influence the
comparison. In particular, it is worth considering potential
systematic errors in the evolution inferred from our dynamical
modeling approach. We discuss two possible effects, but conclude in
the end that neither is likely to be a significant contributor to the
trend seen in Figure~\ref{f:DynFP}.

\subsubsection{Environment}
\label{sss:environment}

It is possible that $M/L$ depends not only on $\sigma_{\rm eff}$, but
also on environment. Our sample of intermediate redshift galaxies
consists of galaxies in rich cluster environments. To study $M/L$
evolution one compares to local galaxies which generally reside in
somewhat different environments. So any local dependence of $M/L$ on
environment would bias the inferred $M/L$ evolution. If this is in
fact an issue, then it is more likely to affect the evolution inferred
from dynamical models and the $M/L$--$\sigma$ relation, than it is to
affect FP studies. The latter generally use Coma as a local comparison
sample. While Coma is not as rich as clusters studied at higher
redshift, it is nonetheless a dense cluster environment. By contrast,
the local comparison sample used to construct the $M/L$--$\sigma$
relation (Table~\ref{t:local}) contains an inhomogeneous mixture of
galaxies in cluster, group and field environments. It has been
suggested that field galaxies might typically be younger, and thus
have lower $M/L$ values, than cluster galaxies (e.g., Diaferio \etal
2001; Bernardi \etal 2003; Annibali \etal 2006). If this were true, and in
particular for galaxies of low $\sigma$ or mass, then this might
explain part of the trend seen in Figure~\ref{f:DynFP}. However,
there is little evidence that any dependence on environment would in
fact be larger for galaxies of low $\sigma$ or mass (Clemens \etal
2006). Also, a significant fraction (25/60) of the galaxies in our
local comparison sample (Table~\ref{t:local}) come from the work of
C06. They explicitly studied the residuals of the $M/L$--$\sigma$
relation as a function of environment, and did not find any
dependence.

To test directly for environmental effects we compared the
$M/L$--$\sigma$ relation for our local comparison sample to the
$(M/L)_{\rm vir}$--$\sigma$ relation for Coma. The zeropoint $Z_{\rm
coma}$ and slope $S_{\rm coma}$ of the latter relation are listed in
Section~\ref{ss:strucevol}. The value of $S_{\rm coma}$ is consistent
at the $\sim 1\sigma$ level with the slope $S = 0.992 \pm 0.054$
inferred for our local comparison sample. Also, the value of $Z_{\rm
coma}$ is consistent with the value $Z = 0.896 \pm 0.010$ for the
local comparison sample if $K = 5.40 \pm 0.23$. This value can be
compared to the average $K = 5.09 \pm 0.19$ for the galaxies in our
local comparison sample, which can be determined directly by equating
the dynamical $M/L$ to $(M/L)_{\rm vir}$ for these galaxies. These two
estimates of $K$ are entirely consistent, especially when taking into
account the previously discussed systematic uncertainties in the
relative distance scales between our local comparison sample and the
Coma cluster. For comparison, C06 found that for data in the $I$-band
$K \approx 4.8 \pm 0.1$. In addition to being a useful consistency
check, these results show that there is no evidence for a strong
environmental dependence of the $M/L$--$\sigma$ relation in the local
universe. The trend in Figure~\ref{f:DynFP} therefore cannot be
attributed to environmental influences on the analysis.

\subsubsection{Spatial Resolution}
\label{sss:resolution}

FP and dynamical modeling analyses rely in different manner on
knowledge of the galaxy surface brightness profile. To calculate the
FP position of a galaxy one needs to know only the integrated
luminosity within the effective radius. By contrast, to perform the
dynamical modeling one needs to model the surface brightness profile
down to very small radii, which requires accurate PSF
deconvolution. Any systematic errors introduced by this are likely to
produce a relative difference between $M/L$ evolution inferred from FP
analysis and dynamical modeling. This is likely to affect low-mass
galaxies, which are smaller on average, more than high-mass
galaxies. So potential PSF deconvolution errors in our dynamical
modeling could in principle produce a trend such as that in
Figure~\ref{f:DynFP}. While this issue may be a contributing factor to
the results in Figure~\ref{f:DynFP}, we do not believe that it can be
the full explanation. The galaxies in our sample that provide evidence
for evolution in the FP tilt have $r_{\rm eff}$ in the range
$0.16''$--$0.80''$ (see fig.~5a of vDvdM06). Most of these galaxies
are quite well resolved even with the $0.1''$ pixels of the HST/WFPC2.
Nonetheless, future higher resolution imaging, such as now possible
with the HST Advanced Camera for Surveys (ACS), would certainly
decrease any sensitivity of the dynamical modeling on PSF
deconvolution.

Even if limited spatial resolution were an issue for our models, it
would not explain the entire trend in Figure~\ref{f:DynFP}. After all,
a shallower trend is seen even in Figure~\ref{virial}, which does not
involve any dynamical modeling at all. The hypothesis of errors due to
limited spatial resolution could at best explain the trend in
Figure~\ref{f:rot}. Galaxies with more rotational support tend to have
lower mass (see Paper~I), therefore tend to be smaller on average, and
therefore could in principle be more affected by spatial resolution
issues.

\section{Summary and Discussion}
\label{s:conc}

Many studies in the past decade have addressed the $M/L$ evolution of
early-type galaxies using the FP. This uses only global photometric
and kinematic quantities and is therefore relatively straightforward
to explore. However, FP evolution equals $M/L$ evolution only if many
simplifying assumptions are met, as discussed in
Section~\ref{s:intro}. The validity of these assumptions has remained
poorly verified. It is therefore important to address $M/L$ evolution
more directly by using dynamical models for spatially resolved
photometric and kinematic data. In Paper~I we constructed two-integral
$f(E,L_z)$ models for 25 visually-classified early type (and in most
cases elliptical) galaxies in the intermediate-redshift ($z \approx
0.5$) clusters CL3C295, CL0016+16 and CL1601+42. Fitting of the models
to surface photometry from HST and kinematics from Keck/LRIS yielded
for each galaxy the average rest-frame $B$-band $M/L$ inside the
spectroscopically explored region. The results allow a critical test
of many of the assumptions that have underlied previous studies of FP
evolution.

To study redshift evolution we needed a suitable comparison sample of
$M/L$ values for local early-type galaxies. We therefore compiled a
sample of 60 galaxies in the local Universe for which detailed
dynamical models were previously constructed to fit spatially resolved
kinematical data. Attention was restricted to galaxies in five
specific modeling studies that addressed large samples. All inferred
$M/L$ values were brought to a homogeneous distance scale, using
distances obtained with the SBF method. Galaxies without SBF distances
were excluded from the sample. All $M/L$ values were transformed to
the $B$ band using either measured or estimated broad-band colors. C06
found that $M/L$ correlates tightly with velocity dispersion, $\log
(M/L) = Z + S \log(\sigma_{\rm eff}/[200 \kms])$, where $\sigma_{\rm
eff}$ is the velocity dispersion inside an aperture of size equal to
the effective radius $r_{\rm eff}$. This refined previous work which
had found that $M/L$ correlates (more loosely) with galaxy luminosity
or mass. We confirm the finding of C06. Our larger, homogenized sample
gives $Z = 0.896 \pm 0.010$ and $S = 0.992 \pm 0.054$. We estimate the
systematic uncertainty in $Z$ due to modeling uncertainties to be
$\Delta Z = \pm 0.02$, based on a comparison of $M/L$ results obtained
from different dynamical modeling studies. The slope that we derive
for the $B$ band differs from the best-fit slope $S = 0.82 \pm 0.06$
found by C06 for the $I$ band. This is due to the well-known fact that
galaxies of low dispersion (or mass) tend to be bluer than those of
high dispersion.

The $M/L$ values inferred for the intermediate-redshift cluster sample
follow a similar relation with $\sigma_{\rm eff}$ as found for the
local galaxies. However, the zeropoint $Z = 0.657 \pm 0.022$ is
smaller than for the local sample. The measured change of zeropoint
with redshift implies that $\Delta \log(M/L) / \Delta z = -0.457 \pm
0.046 \> {\rm (random)} \pm 0.078 \> {\rm (systematic)}$. A comparison
of the FP defined by the high-mass galaxies with $M \gta 10^{11}
\Msun$ in the same sample with that measured for the nearby 
Coma cluster yields $\Delta \log(M/L) / \Delta z = -0.529 \pm 0.049 \>
{\rm (random)} \pm 0.071 \> {\rm (systematic)}$. The systematic
uncertainties in these results are dominated by a variety of issues
that affect our knowledge of the relative distance scale between local
and distant galaxies. Although these systematic uncertainties are
often not explicitly addressed, we stress that they are present in all
studies of $M/L$ evolution that use either dynamical models or the FP.
The results from both methods are consistent with passive evolution of
high-mass galaxies following formation at high redshift, as quantified
in vDvdM06.

Comparison of the $M/L$ evolution inferred from our dynamical modeling
study and from the FP yields excellent agreement for massive galaxies.
This is an important {\it a posteriori} verification of the
assumptions that have underlied all previous FP evolution studies. It
shows that to lowest order FP evolution does indeed measure $M/L$
evolution, as suggested also by the results of Treu \& Koopmans (2004)
for lensing galaxies. It also shows that the subset of hierarchical
structure formation models in which FP zeropoint evolution does not
track $M/L$ evolution (e.g., Almeida \etal 2006) is inconsistent with
our observational understanding of early-type galaxies. Our results
provide no evidence that the galaxies in the sample with the latest
Hubble types (i.e., E/S0, S0/E, S0, or S0/Sb) or the galaxies with the
highest rotation rates tend to have lower $M/L$ values. This might
have been expected if such galaxies transformed recently from
star-forming field galaxies.

While there is broad agreement between the dynamical modeling and the
FP results, we find important differences in the behavior of $M/L$
evolution as a function of $\sigma_{\rm eff}$ (or similarly,
mass). For the dynamically inferred $M/L$--$\sigma$ relation we find
no evidence for a change of the slope with redshift, and therefore no
dependence of $M/L$ evolution on $\sigma_{\rm eff}$. By contrast, our
own work for this same sample and that of many previous authors on
other samples has found that the FP tilt does evolve with redshift.
We studied this difference by analyzing the residuals with respect to
the fitted relations, and by considering also the evolution implied by
$(M/L)_{\rm vir}$ estimates from the virial theorem. Based on this, we
find that the difference between the results from FP evolution and
dynamically determined $M/L$--$\sigma$ evolution can be plausibly
attributed to a combination of two effects: (a) evolution in
structural properties; and (b) the neglect of rotational support in
studies of FP evolution. We investigated other potential explanations
as well, including the possibility that our results may be biased due
to unaccounted local dependencies of $M/L$ on environment, or the
possibility that systematic errors may affect our results for the
smallest galaxies due to the finite spatial resolution of the HST
imaging data. However, we argue that neither of these latter issues
significantly affects our analysis.

Both FP evolution and evolution of the $M/L$--$\sigma$ relation
constrain the $M/L$ evolution of elliptical galaxies. However, these
approaches need not give the same answer (or more generally, the
correct answer) if the structure of galaxies evolves with time. We
find some evidence for this from the fact that the results from FP
evolution and $(M/L)_{\rm vir}$--$\sigma$ evolution differ (in the
sense that there is a slight trend with $\sigma_{\rm eff}$), even when
the analysis is in both cases based on exactly the same global
properties ($\sigma_{\rm eff}$, $r_{\rm eff}$, and $I_{\rm eff}$) and
uses exactly the same local comparison sample (Coma). In general,
neither FP evolution nor evolution of the $M/L$--$\sigma$ relation can
uniquely and correctly determine the amount of $M/L$ evolution, unless
one knows (or correctly assumes) how the structural properties of the
galaxies change with time. This is is not very well constrained
observationally and one therefore has to rely on assumptions and
theoretical insights. Translation of $M/L$--$\sigma$ evolution into
$M/L$ evolution assumes only that the $\sigma_{\rm eff}$ of galaxies
does not change over time. By contrast, translation of FP evolution
into $M/L$ evolution assumes that none of the quantities $\sigma_{\rm
eff}$, $r_{\rm eff}$, $\Sigma_{\rm eff} \equiv I_{\rm eff} (M/L)$, or
the structure constant $K$ change over time. Although elliptical
galaxies are collisionless systems, all these structural quantities
can in fact change through mergers. However, the inferred $M/L$
evolution is in practice used primarily to estimate the mean age of
the stars under the assumption that only the luminosity is evolving
(e.g., vDvdM06). This relaxes the underlying assumptions in the sense
that the correct age is obtained (but not the correct $M/L$ evolution)
as long as any evolution of structural parameters moves galaxies only
along the $M/L$--$\sigma$ relation or along the edge-on projection of
the FP. Models have suggested that structural changes induced by
mergers do indeed approximately have this property (Gonzalez-Garcia \&
van Albada 2003; Boylan-Kolchin, Ma, \& Quataert 2006; Robertson \etal
2006). 

An approach that is based on dynamical models has less built-in
assumptions than approaches (such as those using the FP or $(M/L)_{\rm
vir}$ values) that are based entirely on global or characteristic
quantities. In particular, the $M/L$ values that we determined in
Paper~I account for several known non-homologies between galaxies,
such as differences in axial ratio, brightness profile, rotational
support, and internal velocity distribution. Moreover, kinematical
profiles as a function of radius were calculated and fitted, instead
of just a single characteristic dispersion. We found that the
difference in the $M/L$ evolution inferred from either dynamical
models or the virial theorem correlates with the galaxy rotation rate
$k$. This suggests that the omission of rotation in studies of FP
evolution may be an important oversimplification. Nonetheless, it is
important to stress that many caveats remain even with the dynamical
modeling approach explored here. For example, if the triaxiality or
velocity dispersion anisotropy of early-type galaxies evolves with
redshift then this might bias the results of our dynamical models (as
it would for FP studies). Also, both FP studies and the present work
may suffer from ``progenitor bias''. This is the bias introduced by
the fact that some of today's early-type galaxies may not be
identified as early-type galaxies in samples at higher redshifts (see
vDvdM06 for a discussion of the size of this bias).

In vDvdM06 we reported a steeping of the FP tilt with redshift for our
sample galaxies. This is generally interpreted to mean that low-mass
galaxies have undergone more $M/L$ evolution than high-mass galaxies,
and are therefore younger. However, the results presented here show
that this conclusion need not necessarily be correct: the dynamical
models provide little evidence for a difference in $M/L$ evolution
between low-mass and high-mass galaxies; and the steepening of the FP
tilt may be affected by other issues than $M/L$ evolution. This does
not rule out the possibility that low-mass galaxies have younger
population ages than high-mass galaxies. But it does mean than one
should be careful in drawing conclusions of this nature entirely on
the basis of FP data. In general it is important to test for
differences in population age also on the basis of other
considerations. For example, van der Wel \etal (2005) and Treu \etal
(2005b) considered broad-band colors and spectroscopic diagnostics in
combination with FP residuals to argue for age differences between
field galaxies of different mass. Their results are not necessarily
inconsistent with those presented here, because the situation may be
different for galaxies in the field and in clusters and it may be
different at different redshifts. 

Our work shows that dynamical modeling of large samples of galaxies at
intermediate redshifts provides a powerful new method for the study of
galaxy evolution. It therefore seems useful to expand the present work
to other samples that explore a wider range of redshifts,
environments, and galaxy types.


\acknowledgments We thank Michele Cappellari, Arjen van der Wel and Marijn 
Franx for useful discussions and suggestions.  Part of this research
was carried out at the Kavli Institute for Theoretical Physics in
Santa Barbara, supported in part by the National Science Foundation
under Grant No.~PHY99-07949.




\ifemulate\else\ifemulate\else
\clearpage
\fi\fi


\ifsubmode\else\ifemulate\else
\baselineskip=10pt
\fi\fi



\ifemulate\else
\clearpage
\fi


\ifsubmode\else\ifemulate\else
\baselineskip=14pt
\fi\fi


\ifsubmode
\figcaption{\figcaplocal}
\figcaption{\figcapMLevol}
\figcaption{\figcapMLres}
\figcaption{\figcapDynFP}
\figcaption{\figcapvirial}
\figcaption{\figcaprot}
\figcaption{\figcapMLrotresids}
\clearpage
\else\printfigtrue\fi

\ifprintfig
\ifemulate\else


\clearpage
\begin{figure}
\epsfxsize=0.8\hsize
\centerline{\epsfbox{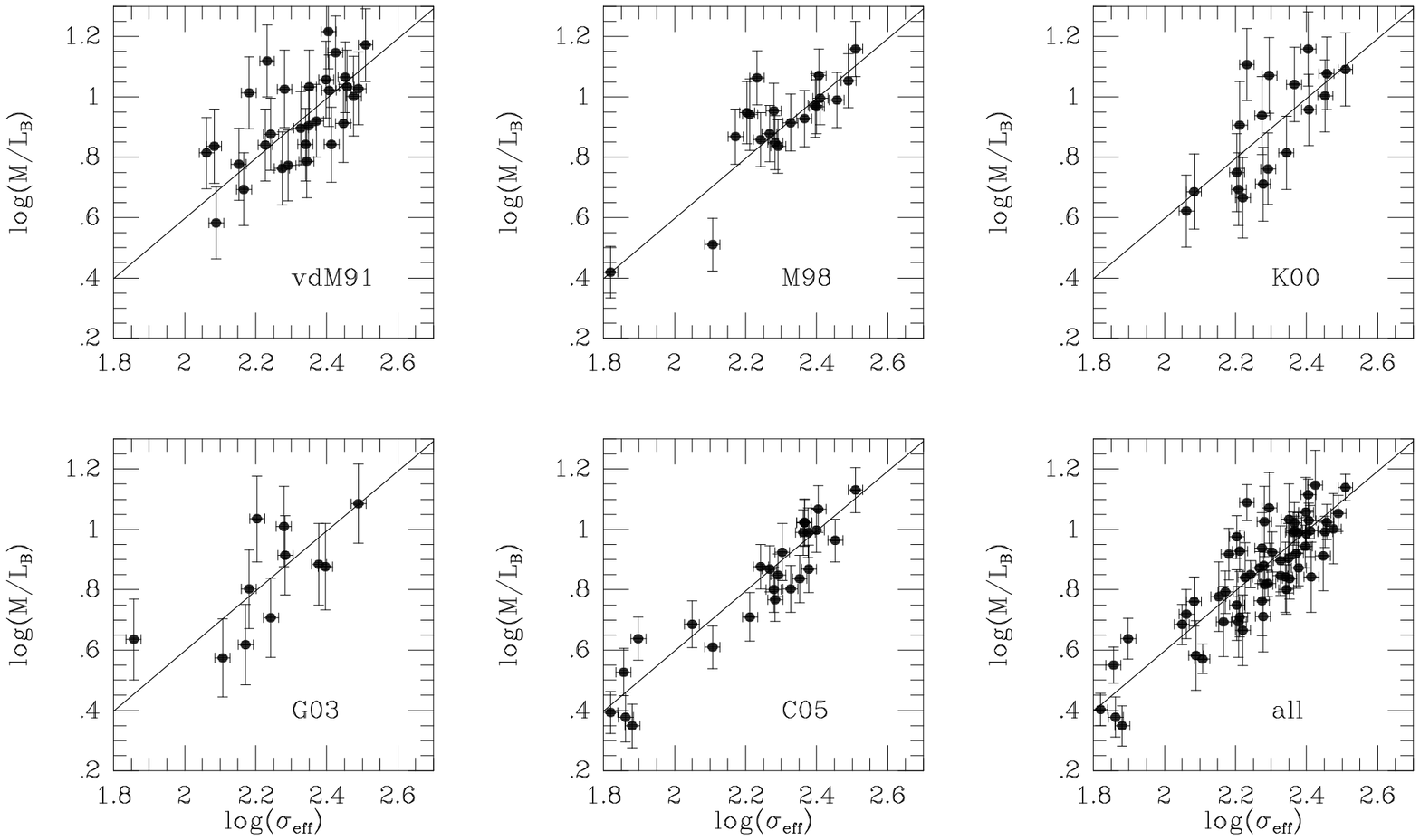}}
\ifsubmode
\vskip3.0truecm
\setcounter{figure}{0}
\addtocounter{figure}{1}
\centerline{Figure~\thefigure}
\else
\figcaption{\figcaplocal}
\fi
\end{figure}


\clearpage
\begin{figure}
\epsfxsize=0.8\hsize
\centerline{\epsfbox{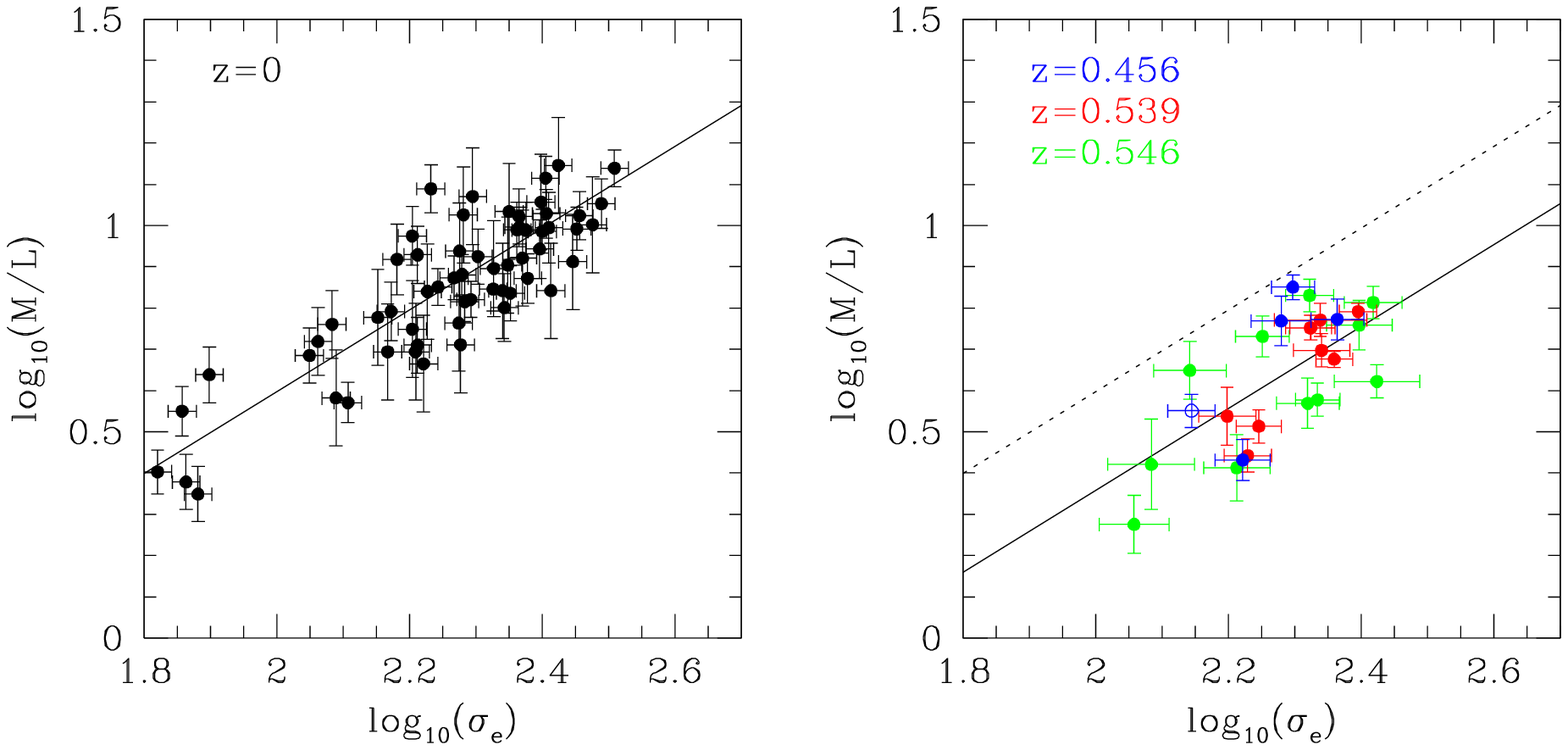}}
\ifsubmode
\vskip3.0truecm
\addtocounter{figure}{1}
\centerline{Figure~\thefigure}
\else
\figcaption{\figcapMLevol}
\fi
\end{figure}


\clearpage
\begin{figure}
\epsfxsize=0.4\hsize
\centerline{\epsfbox{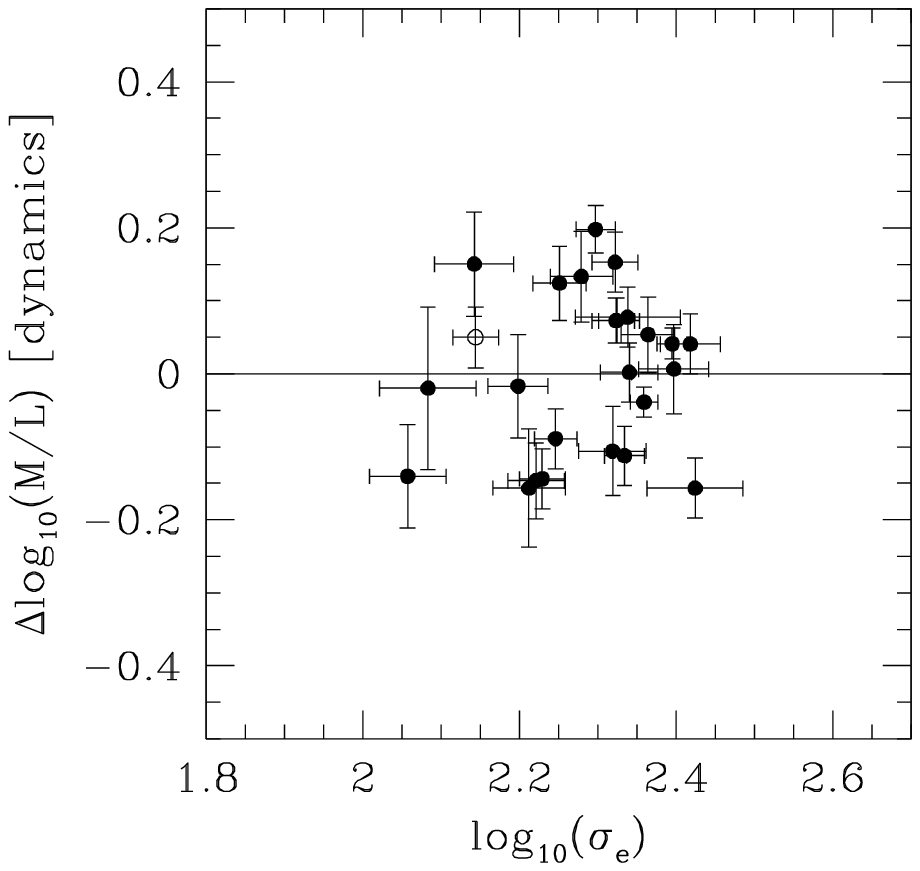}}
\ifsubmode
\vskip3.0truecm
\addtocounter{figure}{1}
\centerline{Figure~\thefigure}
\else
\figcaption{\figcapMLres}
\fi
\end{figure}


\clearpage
\begin{figure}
\epsfxsize=0.4\hsize
\centerline{\epsfbox{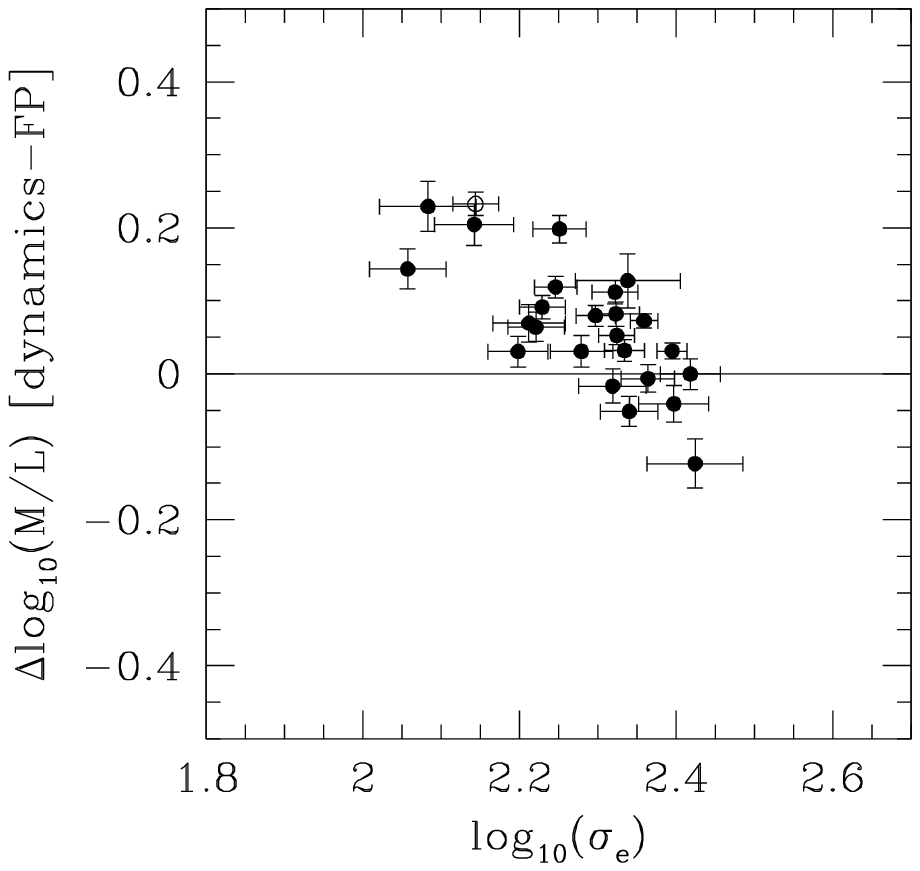}}
\ifsubmode
\vskip3.0truecm
\addtocounter{figure}{1}
\centerline{Figure~\thefigure}
\else
\figcaption{\figcapDynFP}
\fi
\end{figure}


\clearpage
\begin{figure}
\epsfxsize=0.4\hsize
\centerline{\epsfbox{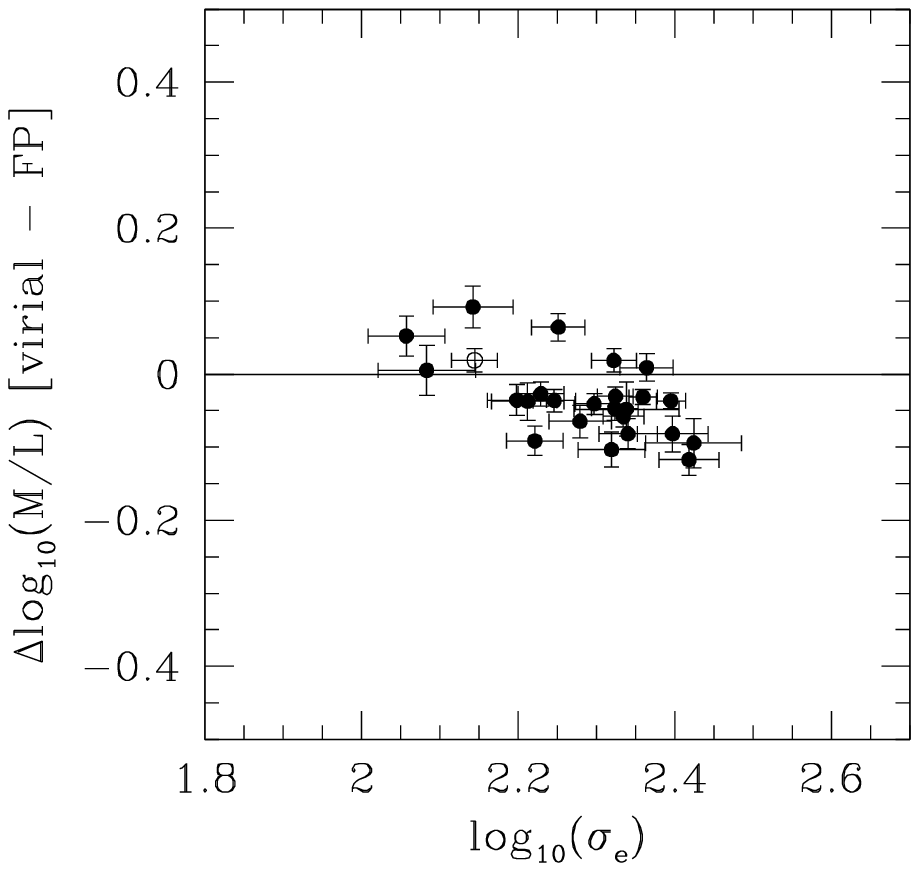}}
\ifsubmode
\vskip3.0truecm
\addtocounter{figure}{1}
\centerline{Figure~\thefigure}
\else
\figcaption{\figcapvirial}
\fi
\end{figure}


\clearpage
\begin{figure}
\epsfxsize=0.4\hsize
\centerline{\epsfbox{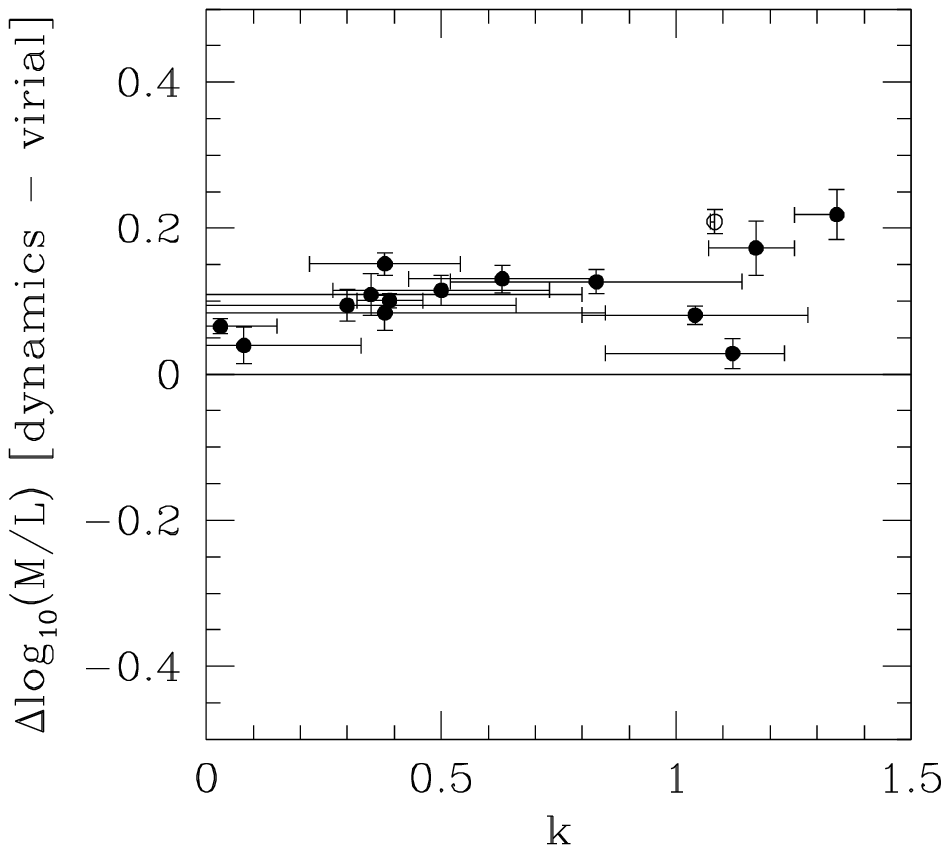}}
\ifsubmode
\vskip3.0truecm
\addtocounter{figure}{1}
\centerline{Figure~\thefigure}
\else
\figcaption{\figcaprot}
\fi
\end{figure}


\clearpage
\begin{figure}
\epsfxsize=0.8\hsize
\centerline{\epsfbox{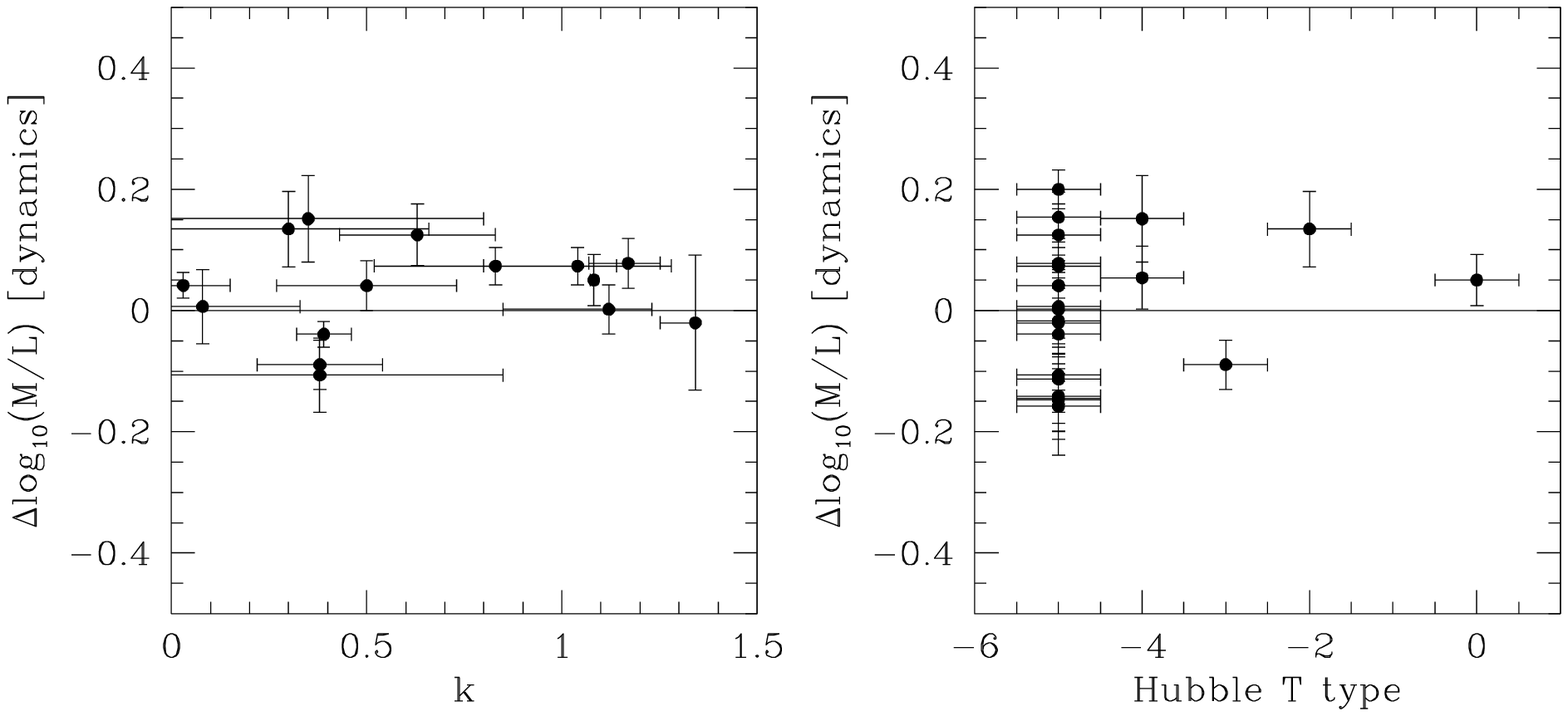}}
\ifsubmode
\vskip3.0truecm
\addtocounter{figure}{1}
\centerline{Figure~\thefigure}
\else
\figcaption{\figcapMLrotresids}
\fi
\end{figure}


\fi\fi


\ifemulate\else
\clearpage
\fi


\ifsubmode\pagestyle{empty}\fi


\ifemulate\else
\begin{deluxetable}{lclc}
\tabletypesize{\footnotesize}
\tablecontentssyserrors
\tablecomments{\tablecommsyserrors}
\end{deluxetable}
\fi


\ifemulate\else
\begin{deluxetable}{lccrcccccc}
\tabletypesize{\small}
\tablecontentslocal
\tablecomments{\tablecommlocal}
\end{deluxetable}
\fi


\ifemulate\else
\begin{deluxetable}{lclc}
\tabletypesize{\footnotesize}
\tablecontentssysFPerrors
\tablecomments{\tablecommsysFPerrors}
\end{deluxetable}
\fi



\end{document}